\documentclass[11pt]{article}
\usepackage{multirow}
\usepackage{cancel}
\usepackage{cite}
\usepackage{amsmath,bbm}
\usepackage{amssymb}
\usepackage{graphicx}
\usepackage[latin1]{inputenc}
\usepackage{mathrsfs}
\usepackage[dvipsnames]{xcolor}
\usepackage[footnotesize]{caption}
\usepackage{epstopdf}
\usepackage{xcolor}
\usepackage{dsfont}
\usepackage{hyperref}

\newcommand{\bea}{\begin{eqnarray}}
\newcommand{\eea}{\end{eqnarray}}
\newcommand{\be}{\begin{equation}}
\newcommand{\ee}{\end{equation}}
\newcommand{\ba}{\begin{array}}
\newcommand{\ea}{\end{array}}

\def\gsim{\mathrel{\rlap{\lower4pt\hbox{\hskip1pt$\sim$}}
    \raise1pt\hbox{$>$}}}

\begin{document}

\begin{titlepage}

\vspace*{-15mm}
\vspace*{0.7cm}

\begin{center}

{\Large {\bf Higgs production from sterile neutrinos \\ at future lepton colliders}}\\[8mm]

\vspace*{0.70cm}

Stefan~Antusch$^{\star \dagger}$\footnote{E-mail: \texttt{stefan.antusch@unibas.ch}},
Eros~Cazzato$^\star$\footnote{E-mail: \texttt{e.cazzato@unibas.ch}},
Oliver~Fischer$^\star$\footnote{E-mail: \texttt{oliver.fischer@unibas.ch}},

\end{center}

\vspace*{0.20cm}

\centerline{$^{\star}$ \it Department of Physics, University of Basel,}
\centerline{\it Klingelbergstr.\ 82, CH-4056 Basel, Switzerland}

\vspace*{0.4cm}

\centerline{$^{\dagger}$ \it Max-Planck-Institut f\"ur Physik (Werner-Heisenberg-Institut),}
\centerline{\it F\"ohringer Ring 6, D-80805 M\"unchen, Germany}

\vspace*{1.2cm}

\begin{abstract}
\noindent 
In scenarios with sterile (right-handed) neutrinos that are subject to an approximate ``lepton-number-like'' symmetry, the heavy neutrinos (i.e.\ the mass eigenstates) can have masses around the electroweak scale and couple to the Higgs boson with, in principle, unsuppressed Yukawa couplings while accounting for the smallness of the light neutrinos' masses.
In these scenarios, the on-shell production of heavy neutrinos and their subsequent decays into a light neutrino and a Higgs boson constitutes a hitherto unstudied resonant contribution to the Higgs production mechanism.
We investigate the relevance of this resonant mono-Higgs production mechanism in leptonic collisions, including the present experimental constraints on the neutrino Yukawa couplings, and we determine the sensitivity of future lepton colliders to the heavy neutrinos.
With Monte Carlo event sampling and a simulation of the detector response we find that, at future lepton colliders, neutrino Yukawa couplings below the percent level can lead to observable deviations from the SM and, furthermore, the sensitivity improves with higher center-of-mass energies (for identical integrated luminosities).
\end{abstract}

\end{titlepage}

\newpage

\section{Introduction} 
Neutrino oscillation experiments have provided us with convincing evidence that (at least two of) the neutrinos are massive. More explicitly, for the three (active) neutrinos of the Standard Model (SM), two differences between the squared masses have been observed, i.e.\ $m_2^2 - m_1^2 = 7.54^{+0.26}_{-0.22}\times 10^{-5}$ eV$^2$ and $|m_3^2 - m_1^2| = (2.43 \pm 0.06) \times 10^{-3}$ eV$^{2}$ \cite{Agashe:2014kda}. The values of the masses themselves cannot be measured via neutrino oscillations, but are bounded to lie below about $0.2$ eV from neutrinoless double beta experiments and cosmological constraints, see for instance ref.~\cite{Gariazzo:2015rra} for a recent review. With only the active neutrino degrees of freedom of the SM, contained in the three SU(2)$_\mathrm{L}$-lepton doublets, it is impossible to add a renormalizable term to the SM which accounts for the observed neutrino masses.

However, renormalisable terms for neutrino masses can be introduced when right-handed (i.e.\ sterile) neutrinos are added to the field content of the SM. These sterile neutrinos are singlets under the gauge symmetries of the SM. They can have a so-called Majorana mass term, that involves exclusively the sterile neutrinos, as well as Yukawa couplings to the three active neutrinos from the SU(2)$_\mathrm{L}$-lepton doublets and the Higgs doublet. 

In the simplistic case, for only one active and one sterile neutrino, with a large Majorana mass $M$ and a Yukawa coupling $y$ such that $M \gg y\, v_\mathrm{EW}$, with $v_\mathrm{EW}$ the vacuum expectation value (vev) of the neutral component of the Higgs SU(2)$_L$-doublet, the mass of the light neutrino $m$ is simply given by $m \approx y^2\, v^2_\mathrm{EW}/M$, while the heavy state has a mass $\sim M$. The prospects for observing this type of sterile neutrino at collider experiments are not very promising: In order to explain the small mass of the light neutrinos (below, say, $0.2$ eV), the mass of the heavy state would need to be of the order of the Grand Unification (GUT) scale, or, alternatively, the Yukawa coupling would be tiny, such that the active-sterile mixing would be highly suppressed.
 
In the more realistic case of three active neutrinos and several\footnote{Since two mass differences in the oscillations of the light neutrinos were observed,  at least two sterile neutrinos are required to give mass to at least two of the active neutrinos.} sterile neutrinos, however, the simple relation from above no longer holds and the possible values of the Majorana masses of the sterile neutrinos and the Yukawa couplings have to be reconsidered. In particular, if the theory entails a ``lepton-number-like'' symmetry, sterile neutrinos with masses around the electroweak (EW) scale and unsuppressed (up to ${\cal O}$(1)) Yukawa couplings are theoretically allowed. This scenario has the attractive features, that one does not have to introduce physics (much) above the EW scale -- which avoids an explicit hierarchy problem -- and one also does not have to introduce otherwise unmotivated tiny couplings.
Various models of this type are known in the literature (see e.g.\ \cite{Wyler:1982dd,Mohapatra:1986bd,Shaposhnikov:2006nn,Kersten:2007vk,Gavela:2009cd,Malinsky:2005bi}). One example is the so-called ``inverse seesaw'' \cite{Wyler:1982dd,Mohapatra:1986bd}, where the relation between light neutrino masses and sterile neutrinos masses is given by $m \,\approx \,\epsilon \,y^2 v^2_\mathrm{EW}/M^2$, where $\epsilon$ is a small quantity that parametrizes the breaking of the protective symmetry. With $\epsilon$ controlling the magnitude of the light neutrino masses, for a given $M$ the coupling $y$ can in principle be large.

In this work, we base our studies on a benchmark scenario which captures the essential features of the realistic case, while it remains more general then specific models: the ``symmetry protected seesaw scenario'' (SPSS), that has also been discussed in ref.~\cite{Antusch:2015mia}. 
In this model, one pair of sterile neutrinos with a generic (approximate) protective symmetry is considered, where the two sterile neutrinos have opposite charges. Additional sterile neutrinos may exist, however it is assumed that their effects can be neglected as far as collider phenomenology is concerned. 
The here relevant parameters of the benchmark scenario are given by the mass parameter $M$, that defines the mass for the two heavy neutrinos due to the protective symmetry, and the moduli of the three Yukawa couplings $|y_{\nu_e}|$, $|y_{\nu_\mu}|$ and $|y_{\nu_\tau}|$ (or, equivalently, of the three active-sterile mixing angles, $|\theta_e|$, $|\theta_\mu|$, $|\theta_\tau|$). We focus on values of $M$ around the electroweak (EW) scale, which might be relevant for collider experiments.   
In ref.\ \cite{Antusch:2015mia} the present constraints on the active-sterile mixing for heavy neutrino masses above $\sim 10$ GeV have been calculated. Therein a combination of precision experiments was considered, which includes EW precision tests, lepton-flavour-violating decays at low energies (most strongly constrained by the results from the MEG collaboration \cite{Adam:2013mnn}) or at the $Z$ pole, tests of lepton universality, decays of the Higgs boson and direct searches at the large electron positron collider (LEP) by the collaborations Delphi \cite{Abreu:1996pa}, Opal \cite{Akrawy:1990zq}, Aleph \cite{Decamp:1991uy} and L3 \cite{Adriani:1993gk}. 
Present and future constraints on EW scale sterile neutrinos have also been studied in the references in \cite{ConstraintsSterile}.

The present experimental bounds on the neutrino Yukawa couplings (and active-sterile mixings) are $\sim 5 \times 10^{-2}$ for heavy neutrinos in the considered mass range. 
These bounds allow for effects of the heavy neutrinos, which could be observed at future lepton colliders\footnote{At the LHC these effects are suppressed by the larger QCD backgrounds and the reduced production cross section of the heavy neutrinos, see ref.\ \cite{Basso} for a detailed analysis.}, such as the Future Circular Collider in the lepton mode (FCC-ee), the Circular Electron Positron Collider (CEPC) or the International Linear Collider (ILC).
One of these effects is a production mechanism for the Higgs boson, which has first been considered in ref.\ \cite{Antusch:2015mia}. In this mechanism the Higgs boson originates from the decay of a heavy neutrino, that has been produced on-shell, and is associated with two light neutrinos. This mechanism is referred to as resonant mono-Higgs production.

In this article, we study the mono-Higgs production mechanism in leptonic collisions for the center-of-energies of 240, 350, and 500 GeV. To be explicit, we consider the FCC-ee in the following and we expect the results to be representative for the CEPC and indicative for the ILC. In order to consolidate the previous estimate \cite{Antusch:2015mia} for the sensitivity of the mono-Higgs production cross section at 240 GeV and to supplement the sensitivities at 350 and 500 GeV, we employ Monte Carlo event generators and simulate the detector response.

The paper is organized as follows: 
In section \ref{sec:SPSS} we introduce the symmetry protected seesaw scenario.
Section \ref{sec:theory} contains a detailed description of the individual contributions to the mono-Higgs production mechanism from the SM and the heavy neutrinos.
In the first part of section \ref{sec:analysis} we estimate the event counts and derive a parton-level sensitivity.
In the second part of section \ref{sec:analysis} we extract a realistic sensitivity from a Monte Carlo event sample (including the SM background), including the simulation of the detector response.
We discuss our results and conclude in section \ref{sec:conclusions}.

\section{Sterile neutrinos at the electroweak scale}
\label{sec:SPSS}
As mentioned in the introduction, it is possible to have sterile (right-handed) neutrinos with masses around the electroweak (EW) scale and unsuppressed (up to ${\cal O}$(1)) Yukawa couplings, when a ``lepton-number-like'' symmetry is realized in the theory. 
The relevant features of seesaw models with such a protective symmetry may be represented in a benchmark scenario, which we refer to as the ``symmetry protected seesaw scenario'' (SPSS) (see also \cite{Antusch:2015mia}) in the following. 

\subsection{The symmetry protected seesaw scenario}
In the SPSS, we consider a pair of sterile neutrinos $N_R^I$ $(I=1,2)$ and a suitable ``lepton-number-like'' symmetry where $N_R^1$ ($N_R^2)$ has the same (opposite) charge as the left-handed $SU(2)_L$ doublets $L^\alpha,\,\alpha=e,\mu,\tau$. The masses of the light neutrinos and other (suppressed) lepton-number-violating effects arise, when this symmetry gets slightly broken.\footnote{We remark that especially at the LHC, lepton-number-violating signatures can provide interesting search channels with low SM background, see e.g.\ refs.~\cite{Deppisch:2015qwa,Banerjee:2015gca,Dib:2015oka}. } 
For the discussion of (lepton-number-conserving) signatures at lepton colliders, however, the effects from the small breaking of the protective symmetry will be neglected.

The Lagrangian density of a generic seesaw model with two sterile neutrinos in the symmetric limit is given by
\be
\mathscr{L} \supset \mathscr{L}_\mathrm{SM} -  \overline{N_R^1} M N^{2\,c}_R - y_{\nu_{\alpha}}\overline{N_{R}^1} \widetilde \phi^\dagger \, L^\alpha+\mathrm{H.c.}\;,
\label{eq:lagrange}
\ee
where we omitted the kinetic terms of the sterile neutrinos, $\mathscr{L}_\mathrm{SM}$ contains the usual SM field content and with $L^\alpha$ and $\phi$ being the lepton and Higgs doublets, respectively. The $y_{\nu_{\alpha}}$ are the complex-valued neutrino Yukawa couplings and the sterile neutrino mass parameter $M$ can be chosen real without loss of generality.

Note that the benchmark scenario posits exactly two right-handed neutrinos, which we assume to be dominating the collider phenomenology.
Furthermore, it captures the general features of symmetry protected seesaw scenarios with more than two right-handed neutrinos, provided that the effects of the additional right-handed neutrinos can be neglected.
This can be the case, when the additional sterile neutrino(s) has large masses, or, alternatively, has zero charge under the ``lepton-number-like'' symmetry. In the limit of exact symmetry, the additional sterile neutrino(s) indeed decouples from the other particles, since no Yukawa couplings to the lepton doublets are allowed and they cannot mix with the other sterile states. 

In the SPSS, the mass matrix of the two sterile neutrinos and the neutrino Yukawa matrix take the form 
\be
M_N = 
\frac{1}{2}\,
\left(\begin{array}{cc} 
0 & M \\
M & 0
\end{array}\right) 
\:,\quad\;
Y_\nu = 
\left(\begin{array}{cc} 
y_{\nu_e} & 0 \\
y_{\nu_\mu} & 0 \\
y_{\nu_\tau} & 0
\end{array}\right),
\ee
where the zeroes correspond to the case of the ``lepton-number-like'' symmetry being exactly realised and are replaced with small quantities when the symmetry is slightly broken.

After EW symmetry breaking, we can write the $5 \times 5$ mass matrix of the electrically neutral leptons as:
\be
\mathscr{L}_{\rm mass} = -\frac{1}{2} \left(\begin{array}{c} \overline{\nu^c_{e_L}} \\ \overline{\nu^c_{\mu_L}} \\ \overline{\nu^c_{\tau_L}} \\ \overline{N_R^1} \\ \overline{N_R^2} \end{array}\right)^T\,
\left( \begin{array}{ccccc}  0 & 0 & 0 & m_e & 0 \\ 0 & 0 & 0 & m_\mu & 0 \\ 0 & 0 & 0 & m_\tau & 0 \\ m_e &  m_\mu & m_\tau & 0 & M \\ 0 & 0 & 0 & M & 0 \end{array}\right) \left(\begin{array}{c} \nu_{e_L}\\\nu_{\mu_L}\\\nu_{\tau_L}\\ \left(N_R^1\right)^c\\ \left(N_R^2\right)^c \end{array} \right) +\mathrm{H.c.}\,,
\label{eq:massmatrix}
\ee
with the Dirac masses $m_\alpha = y_{\nu_\alpha} v_\mathrm{EW}/\sqrt{2}$, and with $v_\mathrm{EW}=246.22$ GeV. 
The diagonalisation of the mass matrix in eq.~(\ref{eq:massmatrix}), referred to as ${\cal M}$ in the following, with a unitary matrix $U$, results in
\be
U^T\, {\cal M}\, U = \text{Diag}\left(0,0,0,M,M\right)\,,
\label{eq:diagonalisation}
\ee
where $U$ is identified with the leptonic mixing matrix. In the symmetric limit, the three light neutrinos are massless and the two heavy neutrinos have degenerate mass eigenvalues.
Note that correction of ${\cal O}(\theta^2)$ to the masses of the heavy neutrinos are neglected. Conversely, when the protective symmetry gets slightly broken, non-zero masses for two of the light neutrinos arise, and e.g.\ a third sterile neutrino could be added in order to explain a non-zero mass for the third light neutrino. 
The mixing of the active and sterile neutrinos can be quantified by the mixing angles, defined as
\be
\theta_\alpha = \frac{y_{\nu_\alpha}^{*}}{\sqrt{2}}\frac{v_\mathrm{EW}}{M}\,.
\label{def:thetaa}
\ee
With the leptonic mixing angles we can express the leptonic mixing matrix $U$ in eq.\ \eqref{eq:diagonalisation}, in the limit of exact symmetry, as:
\be 
U = \left(\begin{array}{ccccc} 
{\cal N}_{e1}	& {\cal N}_{e2}	& {\cal N}_{e3}	& - \frac{\mathrm{i}}{\sqrt{2}}\, \theta_e & \frac{1}{\sqrt{2}} \theta_e 	\\ 
{\cal N}_{\mu 1}	& {\cal N}_{\mu 2}  	& {\cal N}_{\mu 3}  	& - \frac{\mathrm{i}}{\sqrt{2}}\theta_\mu & \frac{1}{\sqrt{2}} \theta_\mu  \\
{\cal N}_{\tau 1}	& {\cal N}_{\tau 2} 	& {\cal N}_{\tau 3} 	& - \frac{\mathrm{i}}{\sqrt{2}} \theta_\tau & \frac{1}{\sqrt{2}} \theta_\tau \\  
0	   	& 0		& 0	&  \frac{ \mathrm{i}}{\sqrt{2}} & \frac{1}{\sqrt{2}}\\
-\theta^{*}_e	   	& -\theta^{*}_\mu	& -\theta^{*}_\tau &\frac{-\mathrm{i}}{\sqrt{2}}(1-\tfrac{1}{2}\theta^2) & \frac{1}{\sqrt{2}}(1-\tfrac{1}{2}\theta^2)
\end{array}\right)\,.
\label{eq:mixingmatrix}
\ee
We remark that the leptonic mixing matrix, as shown above, is unitary up to second order in $\theta_\alpha$. The elements of the non-unitary $3\times 3$ submatrix ${\cal N}$, which is the effective mixing matrix of the three active neutrinos, i.e.~the Pontecorvo--Maki--Nakagawa--Sakata (PMNS) matrix, are given as 
\be
{\cal N}_{\alpha i} = (\delta_{\alpha \beta} - \tfrac{1}{2} \theta_{\alpha}\theta_{\beta}^*)\,(U_\ell)_{\beta i}\,,
\label{eq:matrixN}
\ee
with $U_\ell$ being a unitary $3 \times 3$ matrix. Thus, in the limit of exact symmetry, the SPSS introduces seven additional parameters to the theory, the moduli of the neutrino Yukawa couplings ($|y_{\nu_e}|$, $|y_{\nu_\mu}|$, $|y_{\nu_\tau}|$), their respective phase, and the sterile neutrino mass $M$, which can be studied in the context of collider phenomenology.

\subsection{Weak interactions of the light and heavy neutrinos}
Due to the mixing between the active and sterile neutrinos, the light and heavy neutrino mass eigenstates interact with the weak gauge bosons. The gauge interactions can be expressed by the currents of the neutral fermions in the mass basis, that are given by
\begin{subequations}\label{eq:weakcurrentmass}
\begin{eqnarray}\label{eq:chargedweakcurrentmass}
j_\mu^\pm & = & \sum\limits_{i=1}^5 \sum\limits_{\alpha=e,\mu,\tau}\frac{g}{\sqrt{2}} \bar \ell_\alpha\, \gamma_\mu\, P_L\, U_{\alpha i}\, \tilde n_i\, + \text{ H.c.}\,, \\
j_\mu^0 & = & \sum\limits_{i,j=1}^5 \sum\limits_{\alpha=e,\mu,\tau}\frac{g}{2\,c_W} \overline{\tilde n_j}\, U^\dagger_{j\alpha}\, \gamma_\mu\, P_L\, U_{\alpha i}\, \tilde n_i\,, 
\end{eqnarray}
\end{subequations}
where $g$ is the weak coupling constant, $c_W$ is the cosine of the Weinberg angle and $P_L = {1 \over 2}(1-\gamma^5)$ is the left-chiral projection operator, and where the the mass eigenstates $\tilde n_j$ of the active and sterile neutrinos are defined as
\begin{equation}
\tilde n_j = \left(\nu_1,\nu_2,\nu_3,N_4,N_5\right)^T_j = U_{j \alpha}^{\dagger} n_\alpha\,, \qquad n = \left(\nu_{e_L},\nu_{\mu_L},\nu_{\tau_L},(N_R^1)^c,(N_R^2)^c\right)^T\,.
\end{equation}
Moreover, the neutrino mass eigenstates interact with the Higgs boson. The Yukawa terms in the mass basis, expanded up to ${\cal O}(\theta^2)$, can be expressed as
\begin{equation}
\mathscr{L}_{\rm Yukawa} = \frac{\sqrt{2}\,M}{v_\mathrm{EW}} \left[\sum\limits_{i=1}^3 \left(\vartheta_{i4}^* \overline{N_4^c}+ \vartheta_{i5}^*\overline{N^c_5}\right) \phi^0 \nu_i + \sum\limits_{j=4,5} \vartheta_{jj}^* \overline{N^c_j} \phi^0 N_j \right] +\text{ H.c.}\,,
\label{eq:Lykawa}
\end{equation}
with
\begin{equation}
\vartheta_{ij} = \sum_{\alpha=e,\mu,\tau} U^\dag_{i\alpha}U^{}_{\alpha j}\,.
\end{equation}

The partial decay widths of a sterile neutrino into weak gauge bosons and the Higgs boson, if kinematically allowed, are 
\begin{subequations}
\label{eq:decaywidths}
\begin{eqnarray}
\Gamma(N_j \to W^\pm\, \ell^\mp_\alpha) & = & \frac{|\theta_\alpha|^2}{2} \frac{G_F\,M^3}{4\sqrt{2}\pi}\Pi_{(1+1)}(\mu_W)\,, \\
\Gamma(N_j \to Z\,\nu_i) & = & |\vartheta_{ij}|^2 \frac{G_F\,M^3}{4\sqrt{2}\pi}\Pi_{(1+1)}(\mu_Z)\,, \\
\Gamma(N_j \to h\,\nu_i) & = & |\vartheta_{ij}|^2 \frac{M^3}{8\,\pi\,v_\mathrm{EW}^2}\left(1-\mu_h^{2}\right)^2\,,
\end{eqnarray}
\end{subequations}
where we introduced $\mu_X = m_X/M$, $G_F$ as the Fermi constant, and the kinematic factor 
\begin{equation}
\Pi_{(1+1)}(\mu_X) = {1\over 2}\left(1-\mu_X^2\right)^2 \left(2+\mu_X^2\right)\,.
\end{equation}
For $M \gg m_h = 125$ GeV, the above partial decay widths result in branching ratios of the heavy neutrinos via $W:Z:H$ like 2:1:1.

\subsection{Input parameters}
\label{sec:inputparameters}
For the determination of the theory parameters, we use the set of input parameters with the highest experimental precision, i.e.\ the mass of the $Z$ boson, the fine structure constant (at the $Z$ pole) and the Fermi constant \cite{Agashe:2014kda}.
We note that the Fermi constant is inferred from the decays of the muon and interpreted in the context of the SM, such that we denote it by $G_F^{\rm SM}$ in the following.
\medskip
\begin{center}
\def\arraystretch{1.3}
\begin{tabular}{|l|c|c|c|}
\hline
Input parameter & $m_Z$ [GeV] & $\alpha(m_Z)^{-1}$ & $G_F^{\rm SM}$ [GeV$^{-2}$]\\
\hline
Value & 91.1875(21)  & 127.944(14)  & 1.1663787(6)$\times 10^{-5}$ \\
\hline
\end{tabular}
\def\arraystretch{1}
\end{center}
\medskip

In order to obtain the Fermi constant in the context of the SPSS, it can be related to $G_F^{\rm SM}$ by comparing the respective theory predictions for the muon decay cross sections. With the definition of the charged current interactions, according to eq.~\eqref{eq:chargedweakcurrentmass}, the cross section in the SPSS, for heavy neutrino masses $M \gg m_\mu$, is given by
\be
\sigma^{\rm SPSS}(\mu^- \to e^- \, \nu \, \bar \nu) =  \left( \cal N^{} N^\dag\right)_{ee}\left( \cal N^{} N^\dag\right)_{\mu\mu} \times \sigma^{\rm SM}(\mu^- \to e^- \, \nu \, \bar \nu)\,,
\label{eq:xsectionrel}
\ee
where the summation over all possible final states is implied.
Due to the non-unitarity of the PMNS matrix from eq.\ \eqref{eq:matrixN}, the first factor on the right-hand side of the above equation is not equal to one, such that the cross sections for muon decay are different in both theories. 
The relation in eq.\ \eqref{eq:xsectionrel} between the cross sections fixes the relation of the Fermi constant in the context of the SPSS and the SM to (see e.g.\ \cite{Antusch:2006vwa,Antusch:2014woa,Antusch:2015mia})
\be
 \left(G_F^{\rm SM}\right)^2 = G_F^2(1-|\theta_e|^2)(1-|\theta_{\mu}|^2)\;.
\label{eq:GF}
\ee
This leads to a modification of the theory prediction for a number of other SM parameters, which will in the remainder of this paper be referred to as ``non-unitarity effects''.
In particular, the theory prediction for the weak mixing angle $\theta_W$ (or, more commonly used, $\sin\theta_W$) at tree level (or in the on-shell scheme at any loop order) can be expressed as
\begin{equation}
s_{W}^2 = \frac{1}{2}\left[1-\sqrt{1-\frac{2\sqrt{2}\alpha \pi}{G_F^{\rm SM} m_Z^2} \sqrt{(1-|\theta_e|^2)(1-|\theta_{\mu}|^2)}}
\right]\,.
\label{eq:seff}
\end{equation}
From the relation $m_Z^2 c_W^2 = m_W^2$, we obtain the modified prediction for the $W$ boson mass. 
Furthermore, the vev of the Higgs boson in the SPSS is given by
\begin{equation}
v_\mathrm{EW} = \frac{1}{\sqrt{\sqrt{2}G_F}} = 246.22\left[1-0.25\left(|\theta_e|^2+|\theta_\mu|^2\right)\right]\,.
\end{equation}
A more detailed discussion and up-to-date constraints on the model parameters can be found in \cite{Antusch:2015mia}.

\section{Mono-Higgs production at future lepton colliders}
\label{sec:theory}
The experimental signature of a resonant peak at the Higgs boson mass in the invariant mass spectrum of its decay products plus significant amount of missing energy, we refer to as mono Higgs. 
In the here considered SPSS the missing energy is due to the light neutrinos escaping detection. We note that the heavy neutrinos decay inside the detector volume for the here considered active-sterile mixings and masses.
In this section we study the effects of sterile neutrinos on the cross section for mono-Higgs production in the context of future lepton colliders, i.e.\ the process
\be
e^+ e^- \,\to\, h \bar \nu \nu \,.
\label{process:ee_nnh}
\ee

Generally, in the SPSS we can split the total cross section for this process into the following three contributions
\be
\sigma_{h\nu\nu} \;=\; \sigma_{h\nu\nu}^{\textrm{SM}} + \sigma_{h\nu\nu}^{\textrm{Non-U}} + \sigma_{h\nu\nu}^{\textrm{Direct}} \,.
\ee
The first contribution is the expression for mono-Higgs production in the SM, with the two main mechanisms for Higgs production given by Higgs strahlung and $WW$ fusion. 
The second contribution contains exclusively the non-unitarity effects which modify the low-energy input parameters as well mixing of the active neutrinos. 
The third contribution includes the direct production of Higgs bosons from the decays of heavy neutrinos. The remainder of this chapter is dedicated to the study of these contributions, up to second order in the active-sterile mixing angles in the context of the considered future lepton collider options.

We focus on the center-of-mass energies $\sqrt{s}=240, 350$ GeV, that are being discussed for the FCC-ee \cite{Gomez-Ceballos:2013zzn} (and the CEPC \cite{Ruan:2014xxa}) to study the properties of the Higgs boson and top quark, respectively. We will also include $\sqrt{s}=$ 500 GeV, which can be reached according to present discussion by the FCC-ee working group, see e.g.\ ref.\cite{Tenchini:2014lma}. We remark, that the here considered center-of-mass energies together with integrated luminosities of order ab$^{-1}$ can also be achieved by the ILC \cite{Baer:2013cma}. However, since the linear colliders are considering polarised beams, we limit the discussion in the following to the circular machines. The relevant machine performance parameters are listed in tab.~\ref{tab:machines}. 
\begin{table}[h]
\def\arraystretch{1.1}
\begin{center}\begin{tabular}{|c|ccc|}
\hline
$\sqrt{s}$	& 240 GeV & 350 GeV & 500 GeV  \\
\hline
Experiments 	& FCC-ee, CEPC & FCC-ee (CEPC, ILC) & FCC-ee (ILC) \\
Luminosity/year & 3.5 ab$^{-1}$ & 1.0 ab$^{-1}$ & 0.3 ab$^{-1}$   \\
years& 3 & 3.5 & 3 \\ 
\hline
\end{tabular}\end{center}
\def\arraystretch{1.0}
\caption{Different center-of-mass energies, with currently discussed target integrated luminosity for the FCC-ee \cite{Gomez-Ceballos:2013zzn,Tenchini:2014lma}, that are also representative to some extent for the CEPC \cite{Ruan:2014xxa}. The ILC is included in parentheses, because it is foreseen to operate with polarised beams, which is not considered in the following.}
\label{tab:machines}
\end{table}

\subsection{Mono-Higgs production in the SM}
\label{sec:monoHiggs-SM}
\begin{figure}
\includegraphics[width=0.49\textwidth]{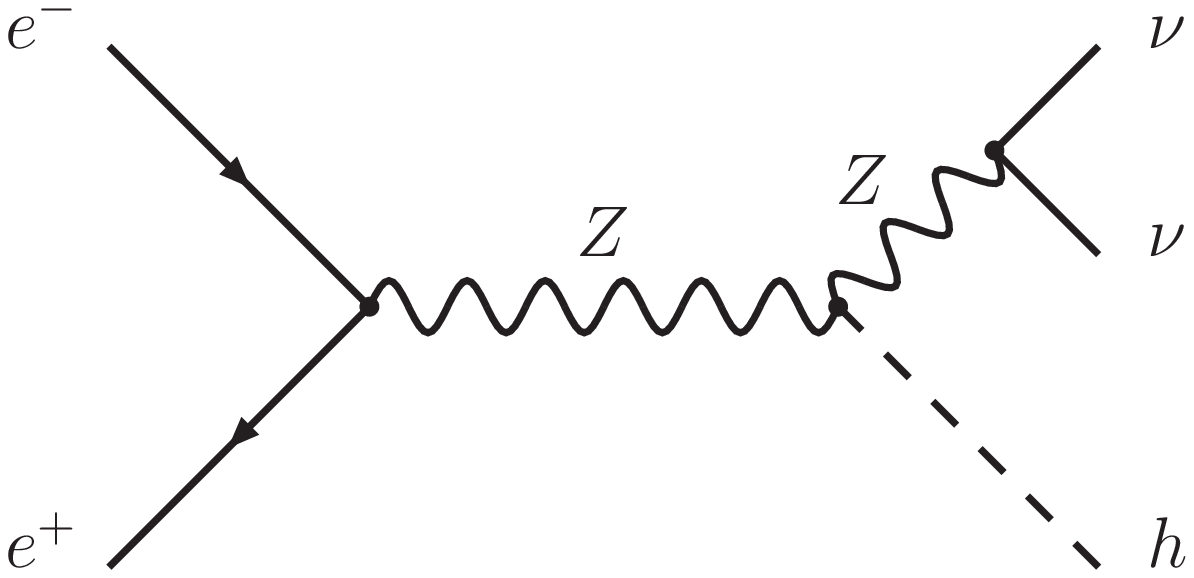}
\includegraphics[width=0.49\textwidth]{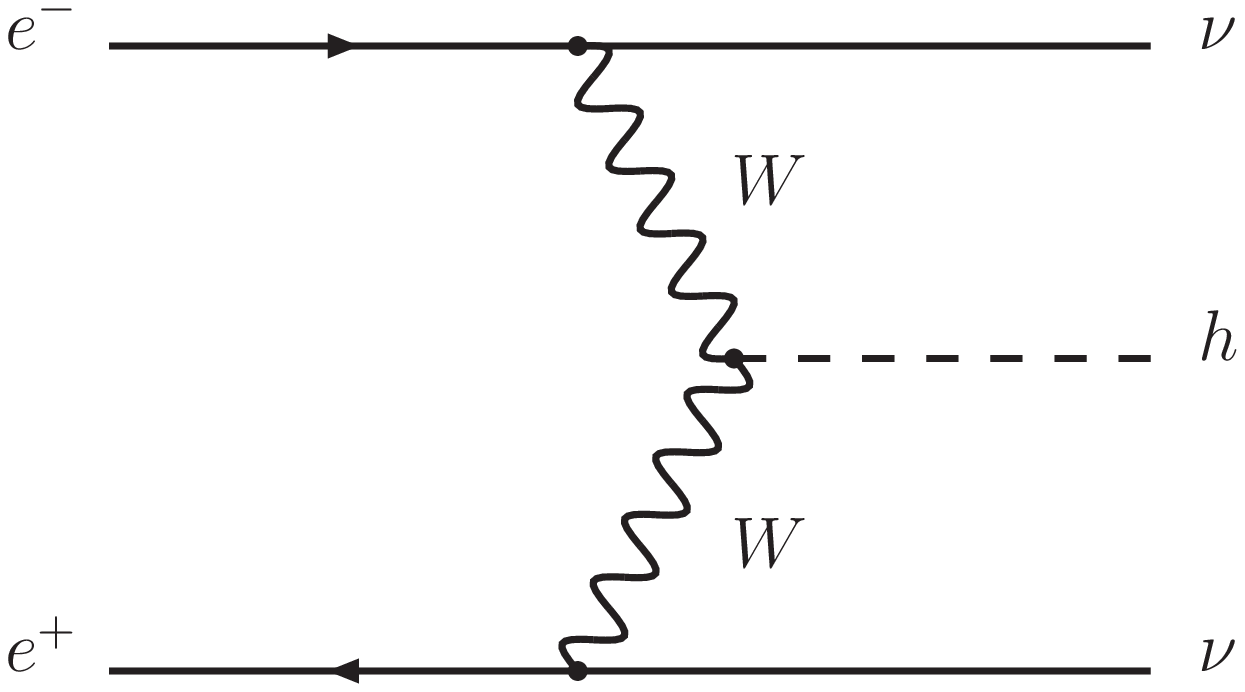}
\caption{The main mechanisms for Higgs boson production plus missing energy in the SM. The Higgs boson is produced by Higgs strahlung or WW fusion.}
\label{fig:higgsproduction}
\end{figure}

At $e^+e^-$ colliders, the most important SM-Higgs-production mechanisms are Higgs strahlung, $e^+e^- \to  Z^* \to Zh$ and $WW$ fusion, $e^+e^- \to h \bar\nu_e \nu_e$, respectively.
The fraction of the $Z$ decays into neutrinos constitutes the Higgs strahlung contribution to the mono-Higgs signature. 
Notice that in $WW$ fusion only electron neutrinos are produced, (since there is no flavour mixing in the SM,) contrary to Higgs strahlung, where all neutrino flavours are produced equally. 
The Feynman diagrams for the two mono-Higgs-production mechanisms are shown in fig.~\ref{fig:higgsproduction}, where we omit the display of explicit indices of the final state neutrinos.

The contribution to the cross section for mono-Higgs production from Higgs strahlung can be expressed in the narrow width approximation as 
\be
\sigma_{h\nu\nu}^{HZ} := \sigma^\textrm{SM}(e^+ e^-\to hZ) \times \text{Br}(Z\to \nu \bar\nu)\,,
\ee
where the branching ratio Br$(Z\to \nu \bar \nu)$ is set to 20.0\% and we implicitly summed over all combinations of final states. The SM Higgs strahlung cross section is given by \cite{Kilian:1995tr}
\be
\sigma^\textrm{SM}(e^+ e^-\to hZ) = \frac{G_f^2 m_Z^4}{24 \pi}\left(v_e^2+a_e^2\right)\lambda^{1 \over 2} \frac{\lambda s+12 m_Z^2}{\left(s-m_Z^2\right)^2}\,,
\label{eq:xHZ}
\ee
with the center-of-mass energy $\sqrt{s}$, the axial- and vector-coupling of the electron-current to the $Z$ boson $a_e = -1/2$ and $v_e = -1/2 + 2 s_W^2$, and the phase-space factor 
\be 
\lambda=\left(1-\frac{(m_h+m_Z)^2}{s}\right)\left(1-\frac{(m_h-m_Z)^2}{s}\right)\,.
\ee

The contribution to the cross section for mono-Higgs production from $WW$ fusion is \cite{Kilian:1995tr}
\be
\sigma_{h\nu\nu}^{WW} := \frac{G_f^3 m_W^4}{4 \sqrt{2} \pi^3} \Pi_{h\nu\nu} \,,
\label{eq:xWW}
\ee
with the phase space factor
\begin{subequations}
\bea
\Pi_{h\nu\nu} & = & \int\limits_{x_h}^1 dx \int\limits_x^1 \frac{dy\, F(x,y)}{(1+(y-x)/x_W)^2}\,, \\
F(x,y) & = & \left( \tfrac{2x}{y^3} - \tfrac{1+3x}{y^2}+\tfrac{2+x}{y}-1\right)\left(\tfrac{z}{1+z} - \log[1+z]\right) + {\tfrac{x}{y^3}} \tfrac{z^2(1-y)}{1+z}\,,
\eea
\end{subequations}
where $x_h = m_h^2/s$, $x_W=m_W^2/s$ and $z=y(x-x_h)/(x\,x_W)$. 

\begin{figure}
\begin{center}
\includegraphics[width=0.70\textwidth]{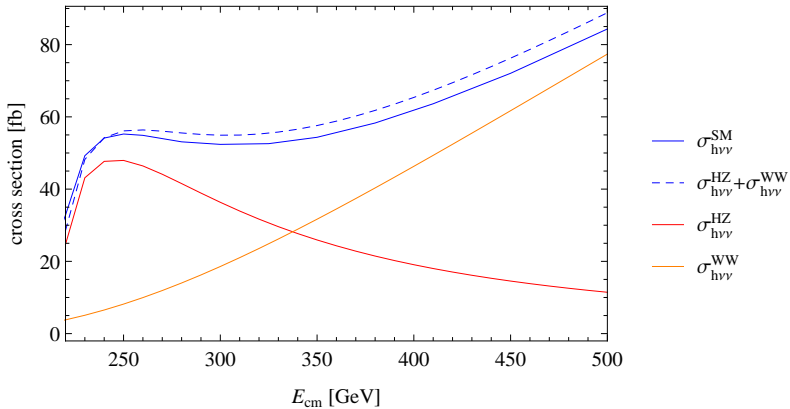}
\end{center}
\caption{
The total mono-Higgs-production cross section and the individual contributions from Higgs strahlung ($\sigma_{h\nu\nu}^{HZ}$) and from $WW$ fusion ($\sigma_{h\nu\nu}^{WW}$), respectively. The dashed blue line denotes the naive sum of $\sigma_{h\nu\nu}^{HZ}$ and $\sigma_{h\nu\nu}^{WW}$, with the small interference term neglected, see text for details.}
\label{fig:vvh_production}
\end{figure}

The mono-Higgs-production cross section $\sigma_{h\nu\nu}^{SM}$ is given by the sum of $\sigma_{h\nu\nu}^{WW},\,\sigma_{h\nu\nu}^{HZ}$ and a contributing interference term. 
We show in fig.~\ref{fig:vvh_production} the individual contributions to the mono-Higgs-production cross section, their naive sum, and the total cross section.
For the sake of simplicity, we neglect the interference term in the following discussion. This is a good approximation since at 240 and 350 GeV it contributes less than $5\%$ to the total cross section. We emphasize, however, that in our analysis we use the full expression for $\sigma_{h\nu\nu}^{SM}$.
As we can see from fig.~\ref{fig:vvh_production}, the cross section for mono-Higgs production at $\sqrt s =240$ GeV is dominated by the contribution from Higgs strahlung, contrary to $\sqrt s = 500$ GeV, where the contribution from $WW$ fusion has taken over. 

\subsection{Non-unitarity effects in mono-Higgs production}
\label{sec:monoHiggs-NU}
In this section, we discuss how the effects from the modified properties of the light (mostly active) neutrinos lead to a deviation of the mono-Higgs-production cross section from the SM prediction. 
This modification manifests itself in the non-unitarity of the effective PMNS matrix $\cal N$ which we refer to as the non-unitarity effects.
Note that these effects do not include exchange, nor production and decay, of the heavy neutrinos.

One part of the non-unitarity effects stem from the modification of the input parameters, as described in section \ref{sec:inputparameters}.
In particular, the dependence of the Fermi constant on the active-sterile mixing parameters, cf.\ eq.~\eqref{eq:GF}, introduces a global change in the definitions for $\sigma_{h\nu\nu}^{HZ}$ and $\sigma_{h\nu\nu}^{WW}$, see eqs.~\eqref{eq:xHZ} and \eqref{eq:xWW}. Also the electroweak parameters sin$\theta_W$ and $m_W$ add to the effect.

The other part of the non-unitarity effects comes from the modification of the vertices according to eq.~\eqref{eq:weakcurrentmass}, where the non-unitary PMNS matrix enters.
The partial mono-Higgs-production cross section $\sigma_{h\nu\nu}^{HZ}$ is therefore proportional to $|\sum_{i,j={1,2,3}} \left( \cal N^\dag N\right)_{ij}|^2$, whereas $\sigma_{h\nu\nu}^{WW}$ is proportional to $|\sum_{i,j={1,2,3}}({\cal N}_{je}^\dag{\cal N}_{ei}^{})|^2$, where the flavour index ``$e$'' is fixed by the incident lepton beams. 
Notice that it is possible to have two different light neutrinos in the final state, i.e.\ $i\neq j$.

We combine the above discussed non-unitarity effects and expand in the small active-sterile mixing parameters to order $\theta^2$, so that we can write the deviation from the SM predicted mono-Higgs-production cross section as:
\be \label{eq:indirecteffect}
\sigma_{h\nu\nu}^{\textrm{Non-U}}
\,=\,
\sigma_{h\nu\nu}^{\textrm{SM}}
   \sum\limits_{\alpha =e,\mu,\tau}
c_\alpha(\sqrt{s})\,|\theta_\alpha|^2 
+{\cal O}(\theta^4)
\,.
\ee
The coefficients $c_\alpha$ are dependent on the center-of-mass energy: firstly, the relative contribution from Higgs strahlung and $WW$ fusion to the total cross section varies with $\sqrt{s}$, and, secondly, both diagrams vary differently with the active-sterile mixing parameters.
In tab.~\ref{tab:cparameters} we list the resulting numerical values of the coefficients for the center-of-mass energies 240, 350 and 500 GeV.

\begin{table}[h]
\def\arraystretch{1.1}
\begin{center}\begin{tabular}{|c|ccc|}
\hline
$\sqrt{s}$/GeV	& 240 & 350 & 500 \\
\hline
$c_{e}$ 	& 0.88 & 0.26 & 0.10\\
$c_{\mu}$ 	& 1.08 & 1.28 & 1.70\\
$c_\tau$ & -0.53 & -0.40 & -0.05 \\
\hline
\end{tabular}\end{center}
\def\arraystretch{1.0}
\caption{List of the coefficients from eq.~\eqref{eq:indirecteffect} obtained with WHIZARD 2.2.7 \cite{Kilian:2007gr,Moretti:2001zz}. The numerical precision of the coefficients is $\pm$ 0.04, $\pm$ 0.05 and $\pm$ 0.05 at the center-of-mass energy of 240, 350 and 500 GeV, respectively.}
\label{tab:cparameters}
\end{table}
For $|\theta_{\tau}|$ substantially smaller than $|\theta_{{e,\mu}}|$, the deviation in the cross section due to non-unitarity is positive, contrary to the case of dominating $|\theta_{\tau}|$, where the negative coefficient $c_\tau$ in eq.~(\ref{eq:indirecteffect}) leads to a negative deviation in the cross section which is formally given by a negative $\sigma_{h\nu\nu}^{\textrm{Non-U}}$.

\subsection{Resonant mono-Higgs production from sterile neutrinos decays}
\label{sec:resonant-monoHiggs}
\begin{figure}
\begin{center}
\includegraphics[width=0.44\textwidth]{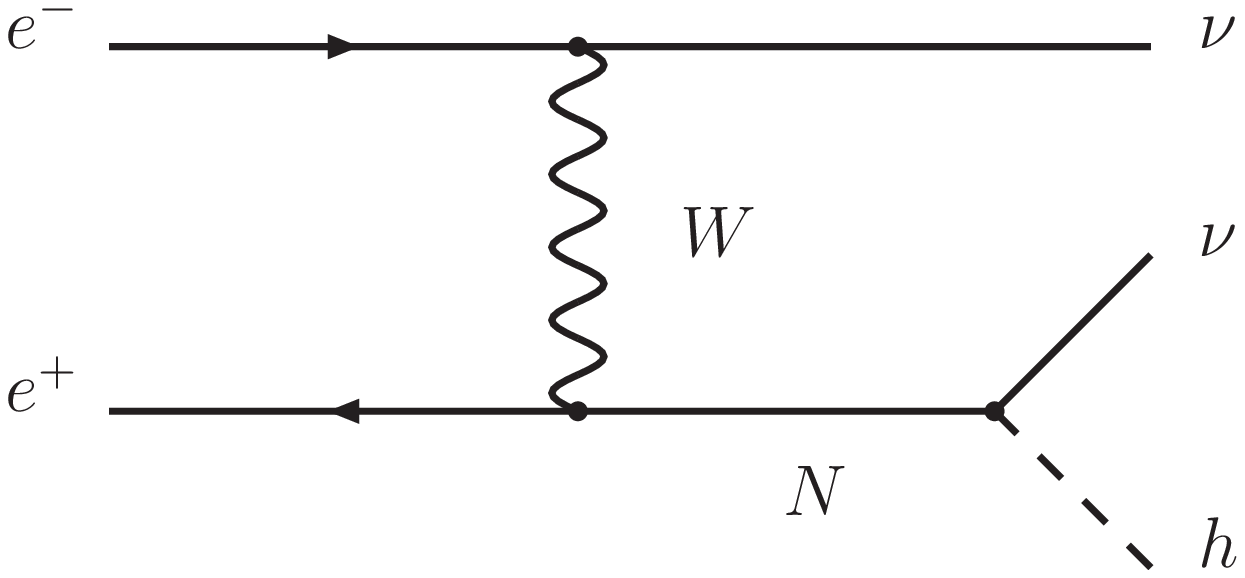}
\includegraphics[width=0.44\textwidth]{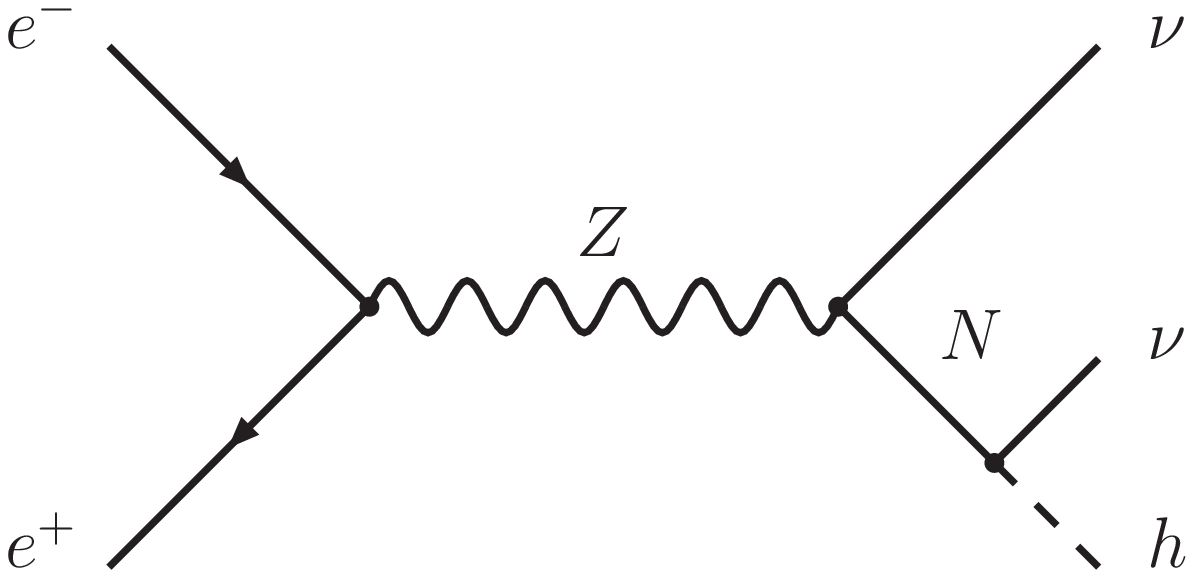}
\end{center}
\caption{The two dominating Feynman diagrams that give rise to the partial mono-Higgs production cross section $\sigma_{h\nu\nu}^\textrm{Direct}$ involving the exchange of heavy neutrinos and leading to a resonant enhancement of the mono-Higgs production.}
\label{fig:vh-production}
\end{figure}

The last contribution to the mono-Higgs-production cross section, $\sigma_{h\nu\nu}^{\textrm{Direct}}$, includes the effects from the exchange of virtual heavy neutrinos, see fig.~\ref{fig:vh-production}, for the corresponding Feynman diagrams. 
We note that the contribution from the s-channel Higgs-exchange diagram to the production of heavy neutrinos is neglected, due to the smallness of the electron Yukawa coupling. This diagram might become relevant when considering muon colliders.
The on-shell production and subsequent decay of a heavy neutrino into a Higgs boson and a light neutrino, yields a resonantly enhanced contribution to the mono-Higgs production.

\begin{figure}
\begin{center}
\begin{minipage}{0.49\textwidth}
\begin{center}\includegraphics[width=0.82\textwidth]{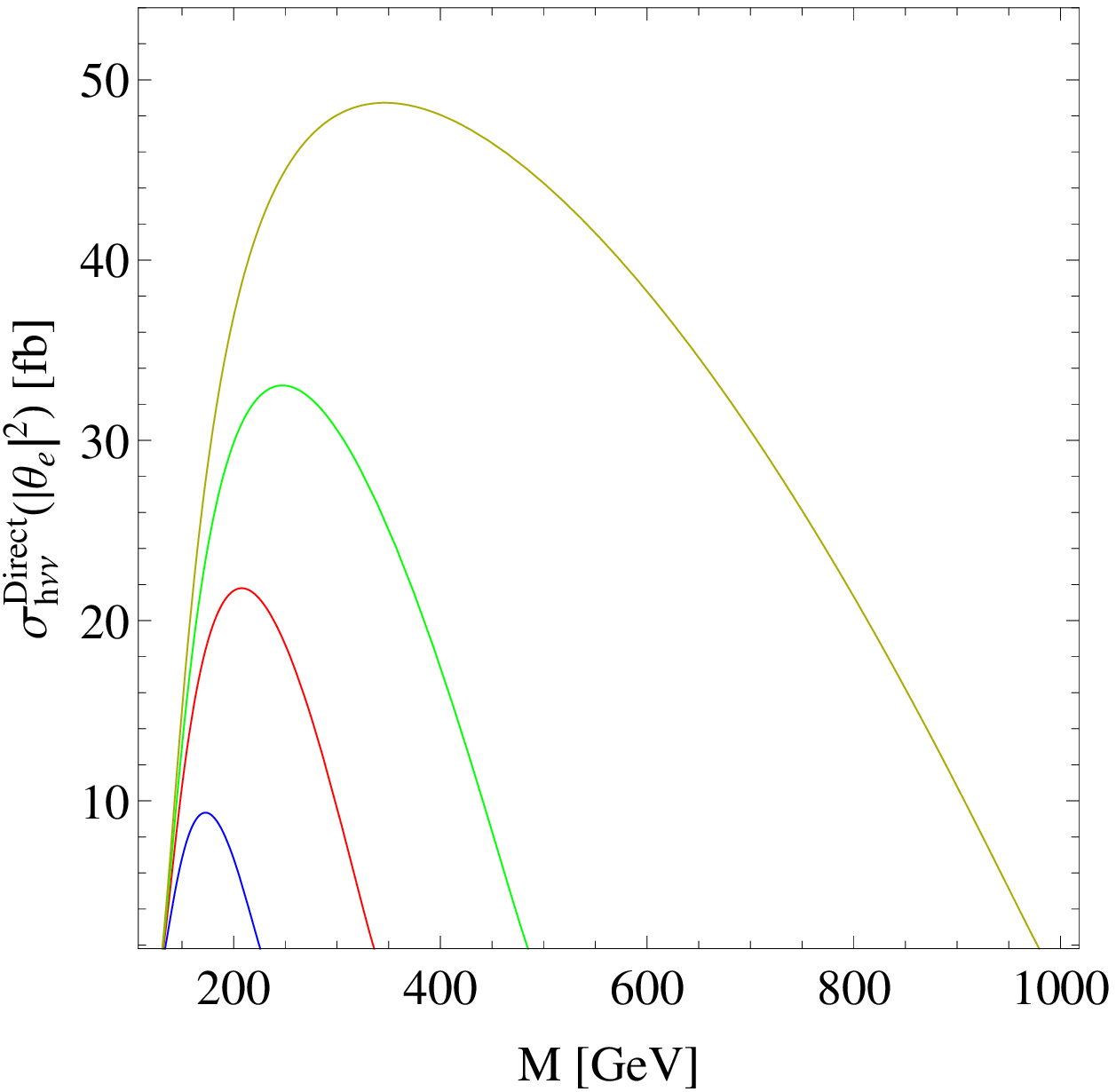}\end{center}
\end{minipage}
\begin{minipage}{0.49\textwidth}
\begin{center}\includegraphics[width=1.1\textwidth]{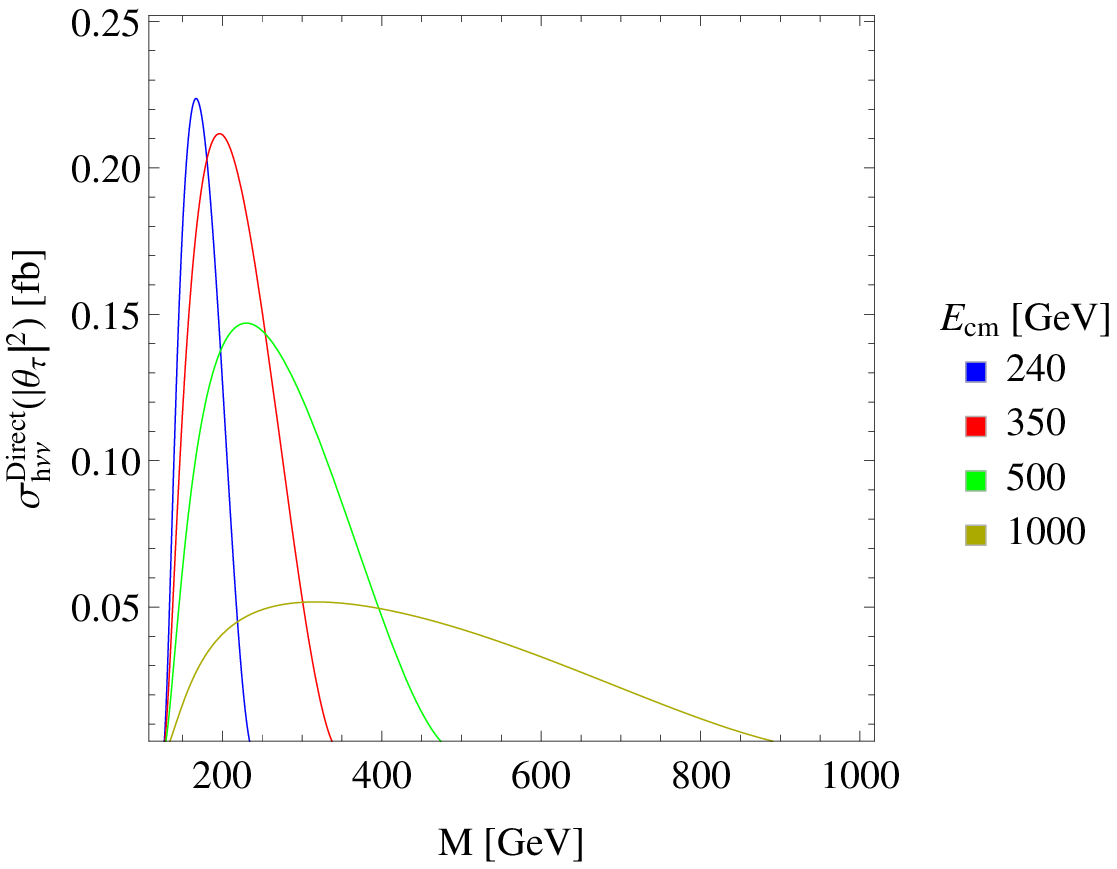}\end{center}
\end{minipage}
\end{center}
\caption{$\sigma_{h\nu\nu}^{\textrm{Direct}}$ as a function of the heavy neutrino mass. {\it Left:} The active-sterile mixing parameter $|\theta_e|^2=0.0018$ is chosen to saturate the 1$\sigma$ upper bound from ref.~\cite{Antusch:2015mia}, while $|\theta_{\mu,\tau}|=0$ are used. {\it Right:} Active-sterile mixing parameter $|\theta_\tau|^2=0.0042$, which saturates the 1$\sigma$ upper bound, while $|\theta_{e,\mu}|=0$.
In this figure, the formula from ref.~\cite{Buchmuller:1991tu} has been used for $\sigma(e^+e^-\to \nu\, N)$.}
\label{fig:sterileproduction}
\end{figure}

The expression  $\sigma_{h\nu\nu}^{\textrm{Direct}}$ also includes the interference between the amplitudes stemming from the Feynman diagrams in fig.~\ref{fig:higgsproduction}, and those from the diagrams in fig.~\ref{fig:vh-production}.
It turns out that the interference of these two sets of amplitudes is negligible, because one part is proportional to the small ratios $\frac{m_e^2}{s}$ and $\frac{m_\nu^2}{s}$, and the other part, resembling the contribution of the Majorana mass of the heavy neutrinos, is cancelled out by the protective symmetry. 
We therefore write to a very good approximation
\be
\sigma_{h\nu\nu}^{\textrm{Direct}} =\sum_{i,j,k} \sigma(e^+ e^- \to N_j\,\nu_i)\times\textrm{Br}(N_j\to h \nu_k)
+ {\cal O}(\theta^4)\,,
\label{eq:sigmanuN}
\ee
with the branching ratios for the heavy neutrinos derived from eqs.~\eqref{eq:decaywidths}, and the production cross section $\sigma(e^+ e^- \to N_j\,\nu_i)$, that can be found for instance in ref.~\cite{Buchmuller:1991tu}.
We show $\sigma_{h\nu\nu}^{\textrm{Direct}}$ as a function of the heavy neutrino mass $M$ for four different center-of-mass energies in the two panels of fig.~\ref{fig:sterileproduction}, using eq.~\eqref{eq:sigmanuN}.
In order to illustrate the effects of the different $\theta_\alpha$, in the left panel of the figure we use the values $|\theta_e|^2=0.0018$, $|\theta_{\mu}|=|\theta_{\tau}|=0$ and in the right panel the values $|\theta_\tau|^2=0.0042$, $|\theta_{e}|=|\theta_{\mu}|=0$. Both sets of example values are within the $1\sigma$ upper bound given in ref.~\cite{Antusch:2014woa}. 

Some remarks on fig.~\ref{fig:sterileproduction} are in order at this point.
The right panel shows the contribution to the production cross section coming exclusively from the s-channel exchange of a $Z$ boson, cf.\ the right diagram in fig.~\ref{fig:vh-production}, which is proportional to $\sum_\alpha |\theta_\alpha|^2$. 
Notice, that the production cross section decreases with increasing center-of-mass energy, since this contribution is suppressed $\sim\frac{1}{s}$ for $s > m_Z$.

The left panel receives contributions from the exchange of both, the $Z$ and the $W$ boson, and for the here considered center-of-mass energies, it is dominated by the latter, cf.\ the left diagram in fig.~\ref{fig:vh-production}. 
Comparing the magnitudes of $\sigma_{h\nu\nu}^{\rm Direct}$ from the left panel with $\sigma_{h\nu\nu}^{\rm SM}$ in fig.~\ref{fig:vvh_production} it is evident that the resonant contribution from heavy neutrinos can indeed be sizeable, and
becomes more relevant at higher energies.
Therefore, for center-of-mass energies of 240 GeV or higher, this results in $\sigma_{h\nu\nu}^{\textrm{Direct}}$ being mostly sensitive to $|\theta_{e}|$, since the vertex of the $W$ boson with the heavy sterile neutrino and the electron is proportional to the matrix element $U_{14}$ or respectively $U_{15}$ of the leptonic mixing matrix, cf.\ eq.~\eqref{eq:mixingmatrix}, with both matrix elements being proportional to $|\theta_{e}|$. 
We remark that a muon collider (see e.g.\ \cite{Neuffer:2015nna}) would allow to test the parameter $| \theta_\mu |$ with great precision and sterile neutrinos with large masses.

\section{Simulation and analysis}
\label{sec:analysis}
In this chapter we quantify the contribution from the decays of sterile neutrinos (cf.\ diagrams in fig.\ \ref{fig:vh-production}), which is considered to constitute the signal for our analysis, over the SM background (corresponding to the diagrams in fig.~\ref{fig:higgsproduction}) through an analysis of Monte Carlo generated event samples. We first analyse the sterile neutrino effects at the parton level, and then include also the simulation of the detector response.
In our analysis we consider processes at the tree level, which is sufficient since in our scenario the one-loop level effects are negligible (cf.\ \cite{Fernandez-Martinez:2015hxa}). Furthermore, we include effects up to order $\theta^2$ in the active-sterile mixing parameters. 
In order to extract the flavour information from the neutrino Yukawa couplings, we use the present constraints from \cite{Antusch:2015mia} and analyse the effect of each one individually.

\vspace{2mm}
{\bf Three cases:} As discussed above, the SPSS has four parameters that are relevant for our considerations: $|y_{\nu_e}|$, $|y_{\nu_\mu}|$, $|y_{\nu_\tau}|$ and the sterile neutrino mass $M$. In the following, to investigate the effects of the Yukawa couplings separately, we will consider the three limiting cases where only one of them is non-zero:
\begin{subequations}\label{eq:cases}
\bea
 \text{ Case I: Effects from $|y_{\nu_e}|$} &\leftrightarrow& y_{\nu_e} \neq 0,\quad  y_{\nu_\mu} = 0 ,\quad  y_{\tau_\mu} = 0\:,\\
 \text{ Case II: Effects from $|y_{\nu_\mu}|$} &\leftrightarrow& y_{\nu_e} = 0,\quad  y_{\nu_\mu} \neq  0 ,\quad  y_{\tau_\mu} = 0 \:,\\
 \text{ Case III: Effects from $|y_{\nu_\tau}|$} &\leftrightarrow& y_{\nu_e} = 0,\quad  y_{\nu_\mu} = 0 ,\quad  y_{\tau_\mu} \neq  0\:.
\eea
\end{subequations}

\vspace{2mm}
{\bf Present constraints on the sterile neutrino parameters:} 
The constraints have recently been calculated in \cite{Antusch:2014woa,Antusch:2015mia}, based on a large set of observables, including e.g.\ the present bounds on EW precision observables, universality test, lepton flavour violating charged lepton decays and the direct searches for neutral leptons at LEP. For heavy neutrino masses in the range $m_Z \leq M \leq \sqrt{s}$, with $\sqrt{s}=240$, 350 and 500 GeV, the constraints can be expressed as upper bounds on the neutrino Yukawa couplings, which, at the $1\sigma$ Bayesian confidence level, can be approximated by:
\be
|y_{\nu_e}| = 0.042 \frac{\sqrt{2}M}{v_\mathrm{EW}},\qquad |y_{\nu_\tau}| = 0.065 \frac{\sqrt{2}M}{v_\mathrm{EW}},\qquad |y_{\nu_\mu}| = 0.015 \frac{\sqrt{2}M}{v_\mathrm{EW}}\,.
\label{eq:upperbounds}
\ee

\subsection{Analysis at the parton level}
As a first step, we consider the contribution from sterile neutrinos to the mono-Higgs production at the parton level. This analysis allows us to establish an order-of-magnitude estimate for the sensitivity of this process to the neutrino Yukawa couplings, and the deviation from the SM prediction.
In order to generate the event distributions at the parton level, we implemented the sterile neutrino (SSPS) benchmark model via Feynrules \cite{Alloul:2013bka} into the Monte Carlo event generator Madgraph5\_aMC$@$NLO \cite{Alwall:2014hca} and analysed the output with madanalysis5 \cite{Conte:2012fm}.

\subsubsection{Definition: signal, background, and significance}
At the parton level, the investigated final state is given by a Higgs boson and two neutrinos, i.e.\ Higgs boson plus missing energy. We define the signal of our analysis to be given by the events that are produced via resonant mono-Higgs production from sterile neutrinos, see section \ref{sec:resonant-monoHiggs}, together with the events stemming from the non-unitarity contribution in section \ref{sec:monoHiggs-NU}. The number of signal events is thus given by $N_S = \left|\sigma_{h\nu\nu} - \sigma_{h\nu\nu}^{\rm SM}\right|\times {\cal L}$
, with the integrated luminosity ${\cal L}$ according tab.\ \ref{tab:machines}. Notice that due to the indirect effect from the input parameters  $\sigma_{h\nu\nu}$ may be smaller than $\sigma_{h\nu\nu}^{\rm SM}$ in case III.
We define the background by the events that stem from Higgs strahlung and $WW$ fusion in the SM (i.e.\ with active-sterile mixing set to zero), as discussed in section \ref{sec:monoHiggs-SM}. The number of SM predicted background events is therefore simply given by 
$N_B = \sigma_{h\nu\nu}^{\rm SM}\times {\cal L}$.

In order to quantify differences between the two models, we define the significance:
\be
\mathcal S= \frac{N_S}{\sqrt{N_B+N_S}}\,.
\label{eq:significance}
\ee
The denominator corresponds to the statistical standard deviation of the total number of events, which is equivalent to the 1$\sigma$ standard deviation when normal distributions are assumed\footnote{The number of events is Poisson distributed which, for the large expected event numbers, approaches the normal distribution.}. 
The above defined significance therefore measures the difference in event counts between the SM and the SPSS in units of standard deviations.

\subsubsection{Number of signal events}
\begin{figure}
\begin{minipage}{0.8\textwidth}
\includegraphics[width=0.5625\textwidth]{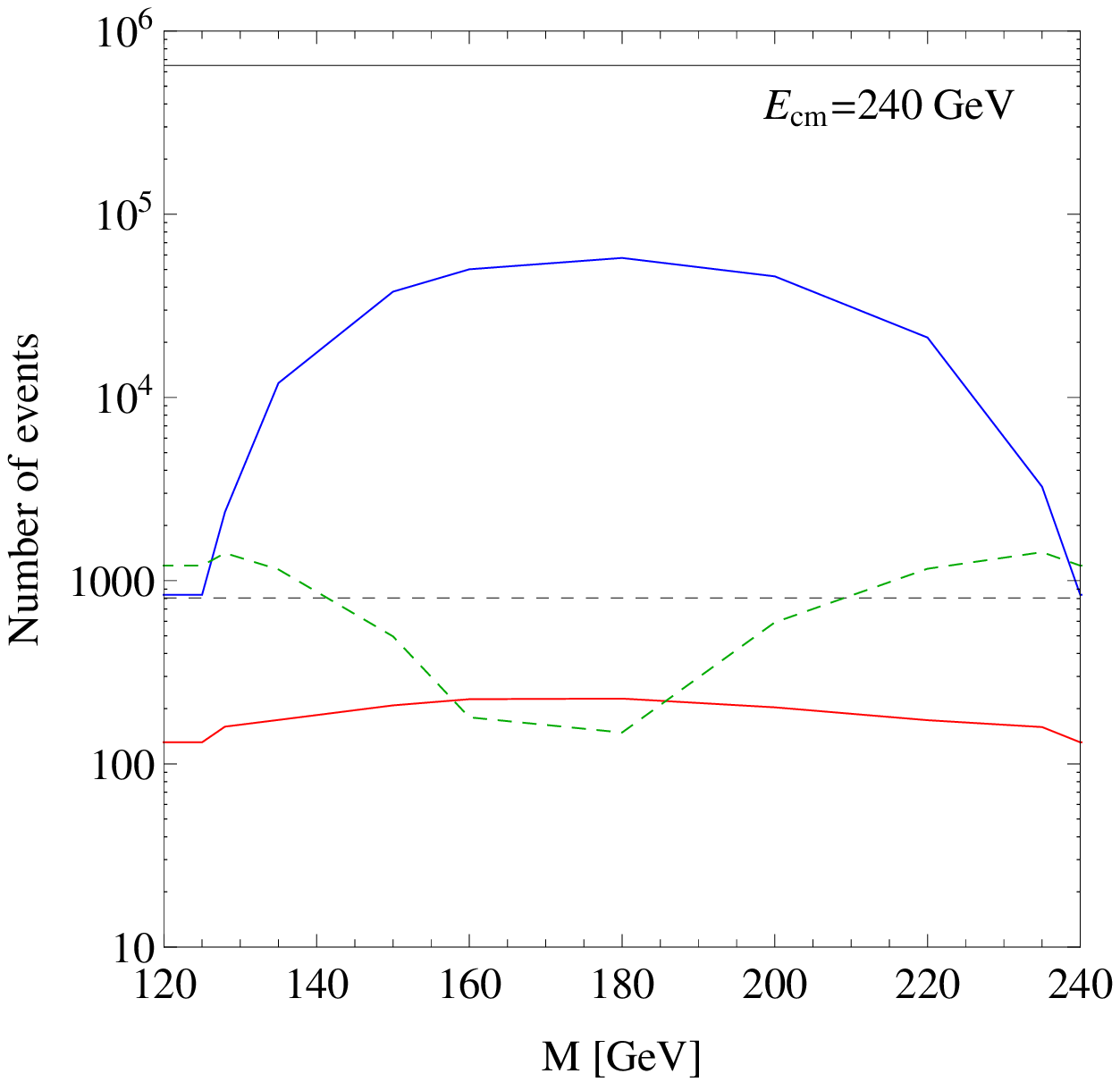}
\end{minipage}
\begin{minipage}{0.15\textwidth}
\mbox{}\hspace{-100pt}
\includegraphics[width=1.3\textwidth]{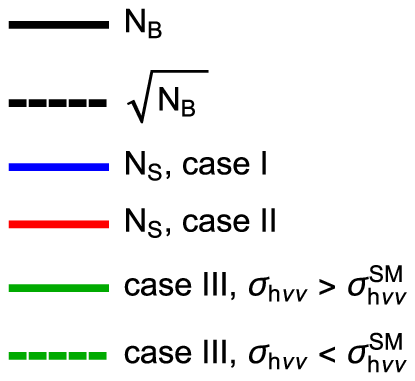}
\end{minipage}

\includegraphics[width=0.45\textwidth]{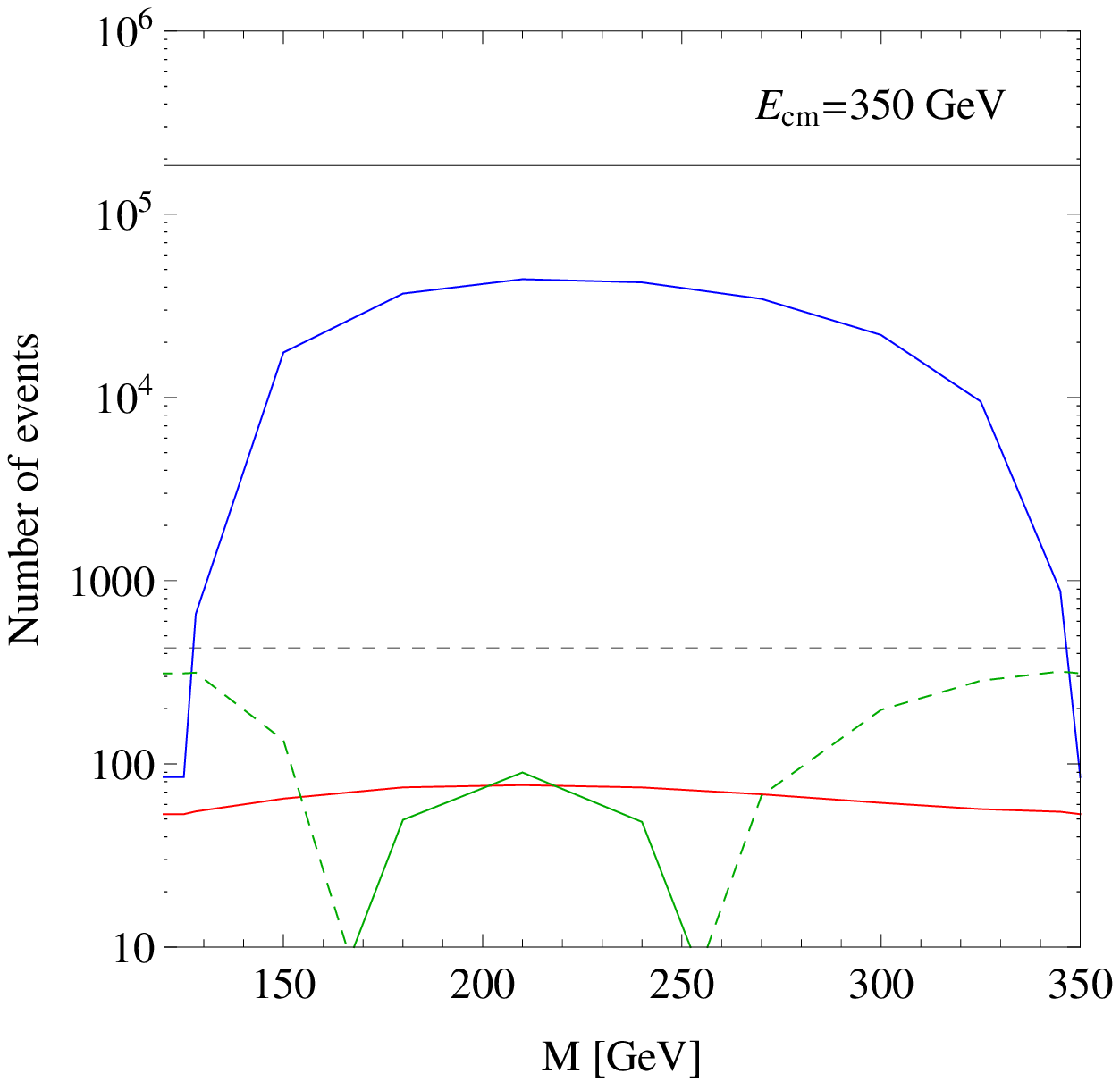}
\includegraphics[width=0.45\textwidth]{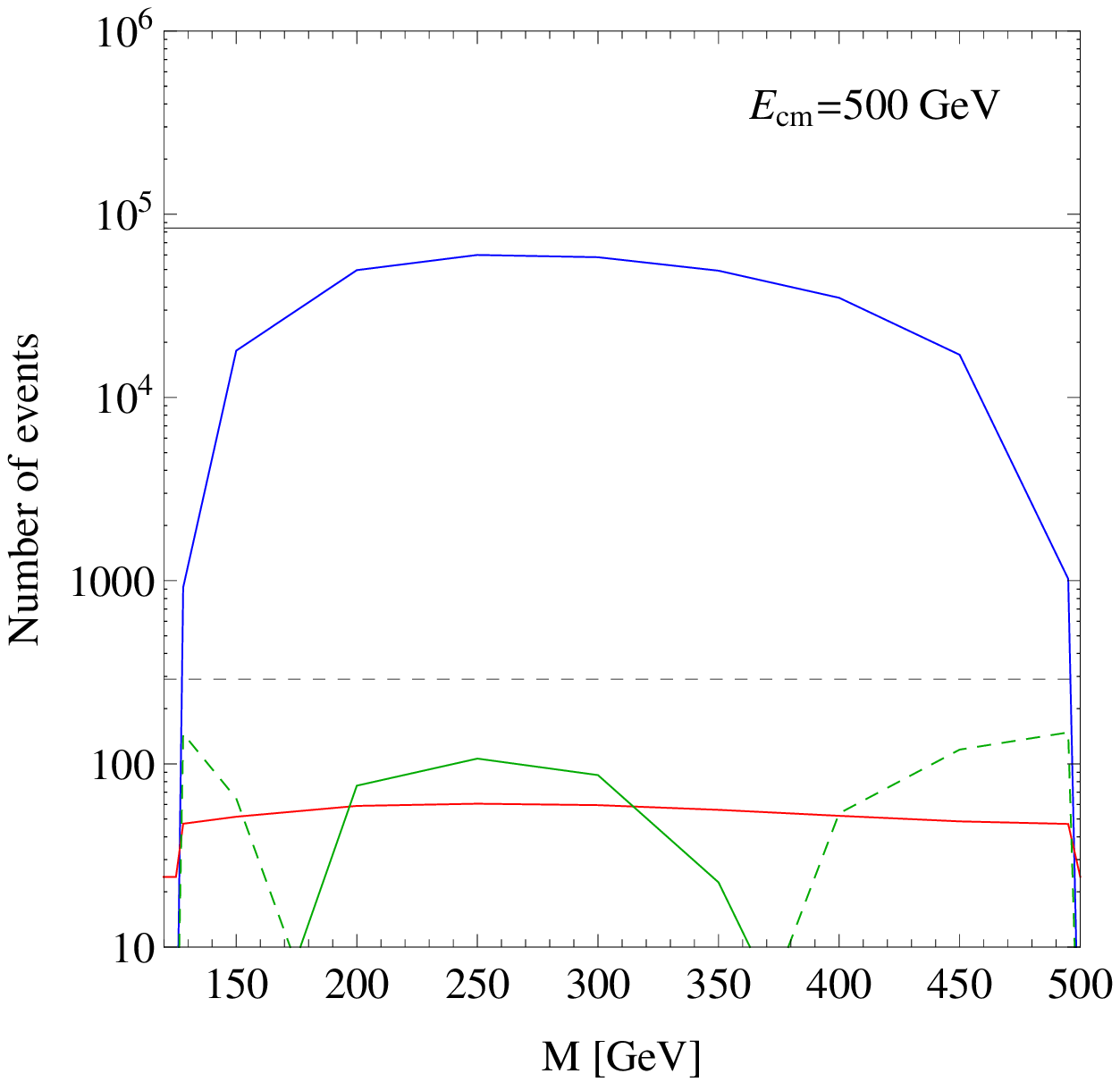}
\caption{Number of signal events $N_S$ in mono-Higgs production at the parton level for the three cases as defined in eq.\ \eqref{eq:cases}, with Yukawa coupling values compatible with present $1\sigma $ bounds from refs.~\cite{Antusch:2014woa,Antusch:2015mia}.
The solid black lines denote the number of background events $N_B$, and the dashed black lines denote $\sqrt{N_B}$.
The machine performance parameters are specified in tab.~\ref{tab:machines}.} 
\label{fig:deltaNmax}
\end{figure}

We estimate the number of mono-Higgs-produced signal events $N_S$, stemming from each flavour corresponding to the cases I,II and III of eq.~({\ref{eq:cases}}), that are compatible with the present upper bounds on $|y_{\nu_e}|$, $|y_{\nu_\mu}|$ and $|y_{\nu_\tau}|$ from eq.~(\ref{eq:upperbounds}) for the FCC-ee.
The number of signal events were calculated from the Madgraph5\_aMC$@$NLO-generated cross sections for eight values of $M$ at 240 GeV and nine values of $M$ for each, 350 and 500 GeV. They typically lead to an excess over the number of background events and are shown, together with the number of background events, in fig.~\ref{fig:deltaNmax}, for the considered center-of-mass energies: 

In the figure, the blue and red lines show the results for the cases I and II, respectively, where the parameters $|y_{\nu_e}|$ and $|y_{\nu_\mu}|$ are non-zero. 
For case III, with non-zero $|y_{\nu_\tau}|$, a partial cancellation between non-unitarity effects and direct production of heavy neutrinos occurs. The situation, where $\sigma_{h\nu\nu} < \sigma_{h\nu\nu}^{\rm SM}$ is shown by the dashed green line, while $\sigma_{h\nu\nu} > \sigma_{h\nu\nu}^{\rm SM}$ is denoted by the solid green line. The number of SM background events is shown as a solid black line. The dashed black line corresponds to $\sqrt{N_B}\approx \sqrt{N_B + N_S}$ for the here considered event numbers.

As fig.\ \ref{fig:deltaNmax} shows, up to ${\cal O}(10^5)$ signal events on the parton level can be produced for the machine parameters from tab.\ \ref{tab:machines}, when non-zero $|y_{\nu_e}|$ compatible with its present bounds is considered. 
Comparing this to the SM predicted number of background events $N_B$ we see that the contribution of the heavy neutrinos to the mono-Higgs-production cross section can be sizeable.
As anticipated in the previous section, $|y_{\nu_e}|$ has by far the largest impact on the mono-Higgs production cross section. 

\subsubsection{Sensitivity to sterile neutrino parameters}
\label{sec:parton-sensitivity}

\begin{figure}
\begin{center}
\begin{minipage}{0.7\textwidth}
\includegraphics[width=1.0\textwidth]{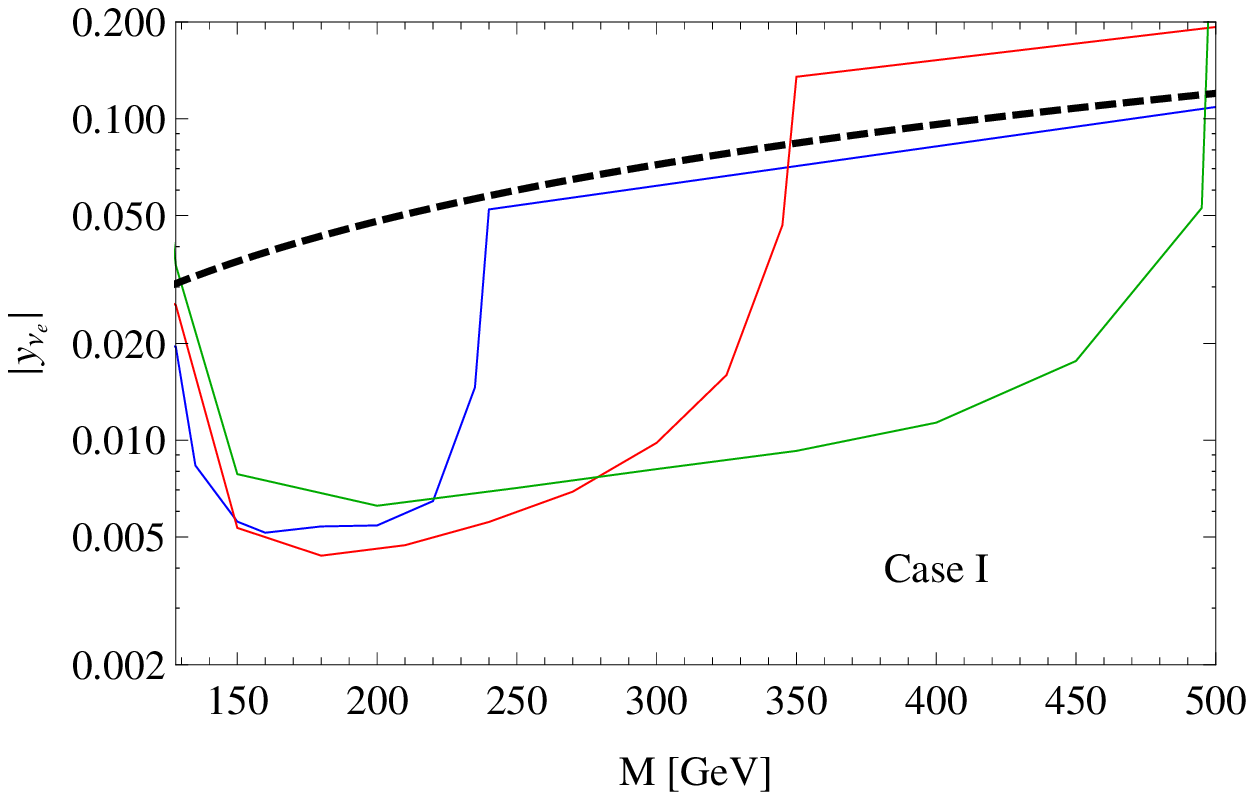}
\end{minipage}
\begin{minipage}{0.29\textwidth}
\centering
\includegraphics[width=0.8\textwidth]{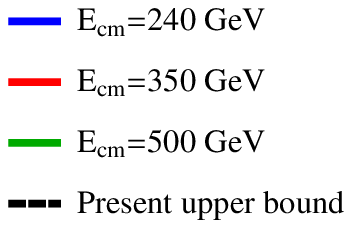}
\end{minipage}
\end{center}
\caption{Sensitivity of the mono-Higgs production cross section to the neutrino Yukawa couplings at the parton level at $1\sigma$, with the machine performance parameters from tab.~\ref{tab:machines}.}
\label{fig:sensitivity}
\end{figure}

We now turn to the possible sensitivity of the mono-Higgs channel at future lepton colliders (cf.\  tab.~\ref{tab:machines}) to the neutrino Yukawa coupling $|y_{\nu_e}|$ (respectively the active-sterile mixing parameter $|\theta_{e}|$) for a given $M$.   
The sensitivity is defined as the value of $|y_{\nu_e}|$ that corresponds to a significance of $\mathcal S=1$ (cf.\ eq.\ \eqref{eq:significance}), i.e.\ to a signal at the 1$\sigma$ level.\footnote{Note that $\mathcal S=1$ corresponds to 84\% confidence level for a one-sided normal distribution, which is chosen here such that the results derived in the following can directly be compared with the corresponding limits in ref.~\cite{Antusch:2015mia}.}
Notice, that the sensitivity to the neutrino Yukawa couplings $y_{\nu_\mu}$ and $y_{\nu_\tau}$ does not improve the present bounds, which is why we omit to discuss the sensitivity for case II and III.
We show the resulting sensitivity to the modulus of the neutrino Yukawa coupling $y_{\nu_e}$ in fig.~\ref{fig:sensitivity}. In the figure, the red, blue and green line corresponds to the sensitivities for $\sqrt{s}=240, 350$ and $500$ GeV, respectively, and the black dashed line denotes the present constraints from \cite{Antusch:2015mia}. 
We have simulated $10^6$ background events and the same number of events for eight values of $M$ for $\sqrt{s}=240$ GeV, and nine values of $M$ for each, $\sqrt{s}=350$ and 500 GeV. For each simulation, we have optimised the cuts to obtain highest sensitivity. 

Remarkably, apart from probing a wider mass range, the sensitivity to $|y_{\nu_e}|$ at 350 GeV is comparable to the center-of-mass energy of 240 GeV, despite the lower integrated luminosity. The same is true for 500 GeV, where 1 ab$^{-1}$ can lead to comparable sensitivities to 350 and 240 GeV even for $M \sim 200$ GeV, which is due to the evolution of $\sigma_{h\nu\nu}^{\rm SM}$ and $\sigma_{h\nu\nu}^{\rm Direct}$ with the center-of-mass energy.
Furthermore it is worth noting that for $M<m_h$ and $\sqrt{s}<M$ the sensitivity stems from non-unitarity effects, i.e.\ the indirect effects from active-sterile mixing on the input parameters and the modified interactions of the light neutrinos. In this case the number of signal events is given by eq.\ \eqref{eq:indirecteffect}. For $\sqrt{s} = 240$ GeV we find that even for $M<m_h$ and $\sqrt{s}<M$, the present constraints on $y_{\nu_e}$ allow for a $\sim 1\sigma$ deviation from the SM prediction for the mono-Higgs-production cross section.

\subsection{Reconstruction with the ILD detector}\label{sec:Analysis}
In this section we describe the relevant SM background and how to extract the mono-Higgs candidates from the reconstructed events after the simulation of the detector response. 
From those mono-Higgs candidate events we calculate a more realistic sensitivity of the mono-Higgs channel to the neutrino Yukawa coupling $y_{\nu_e}$.
Furthermore, we show that the resonant mono-Higgs production can also lead to a contamination of the mono-Higgs candidate event sample, when ``standard cuts'' are applied. 

For the analysis we have generated the signal and background with the Monte Carlo event generator WHIZARD~2.2.7 \cite{Kilian:2007gr,Moretti:2001zz}, which allows for the appropriate simulation of leptonic collisions including initial state radiation. We remark that the effects from beamstrahlung are negligible for the here considered center-of-mass energies and will be neglected in the following.
The parton showering and hadronisation has been carried out with PYTHIA~6.427 \cite{Sjostrand:2006za} and the events were reconstructed with the ILD detector card using Delphes~3.2.0 \cite{Cacciari:2011ma}.

\subsubsection{Signal and background in the mono-Higgs channel}
For the analysis at the reconstructed level, the parton level final states have to be transformed into reconstructed objects. In particular, the light neutrinos manifest themselves as missing energy, and the Higgs bosons decay into $b\bar b\,(57.7\%),$ $WW^*\,(21.6\%),$ $gg\,(8.50\%),$ $\tau^+ \tau^- (6.37\%),$ $c\bar c\, (2.66\%),$ and $ZZ^*$ $(2.46\%)$. Higgs boson candidates can be reconstructed from its decay products, which have an invariant mass around $m_h$. 
In order to obtain better statistics for resonantly produced mono-Higgs events from heavy neutrino decays, we focus on the Higgs decays into two hadronic jets (di-jet) which have a very large combined branching ratio of $\sim$ 70\%. 
The di-jet plus missing energy signature comprises our mono-Higgs search channel such that we select events with two hadronic jets with an invariant mass of $100$ GeV $ \leq M_{jj} \leq 140 $ GeV.

The signal is here given by events that stem from the decays of the heavy neutrinos, which add to the number of events in the search channel.
When considering inclusive processes on the reconstructed level, two mechanisms involving heavy neutrinos contribute to the signal: the resonant mono-Higgs and the resonant mono-Z production mechanisms, where the latter is defined analogously to the former, with the Z originating from the decay of a heavy neutrino.
However, the invariant mass of the resonant mono-Z produced di-jet is $\sim m_Z$, such that the above defined cuts for the mono-Higgs search channel essentially remove this contribution from the signal.
We remark that the resonant mono-Z production constitutes an independent search channel for the heavy neutrinos, and provides an important consistency check for this model, because the relative amount of additional (resonantly produced) events at the Higgs and Z pole, respectively, is predicted by the model parameters. A detailed study of this channel is beyond the scope of this paper and is therefore left for future work.

For the SM background, we include all processes with a four fermion final state, that can be (mis-)identified as a mono-Higgs-candidate event. 
We do not consider processes with di-electrons or di-muons in the final state as background, because it is very unlikely to misidentify two light leptons as a jet at the same time.

The dominating background is given by $q \bar q \,\nu \nu$, with $q=b,\,c,\,g$ stemming from mono-Higgs production in the SM, cf.\ section~\ref{sec:theory}. 
In addition to the mono-Higgs production process, we find the subdominant background consists in processes with $q \bar q \,\nu \nu$ final states, where the quarks $q=b,\,c,\,s,\,d,\,u$ are produced via gauge bosons and in radiative processes.
We note that, due to our selection criterion of the invariant di-jet mass being around $m_h$, most of the backgrounds that stem from gauge boson decays are efficiently suppressed to below the percent level.
Other subdominant backgrounds come from final states with four hadronic jets, $e^+ e^-$ plus di-jet, 2 $\tau$-jets plus di-jet, and single-top and $t \bar t$ final states, when kinematically allowed.

We simulate and reconstruct $10^5$ events for each final state, with the exception for di-b-jet plus missing energy, where 3$\times 10^6$ events have been simulated and reconstructed. 
We note that we simulate inclusive processes such that the interference between the possible production mechanisms is accounted for.
A detailed list of the included backgrounds and their corresponding cross sections is given in tab.~\ref{tab:processtableSM} in the Appendix. 

\begin{figure}
\begin{center}\includegraphics[width=0.6\textwidth]{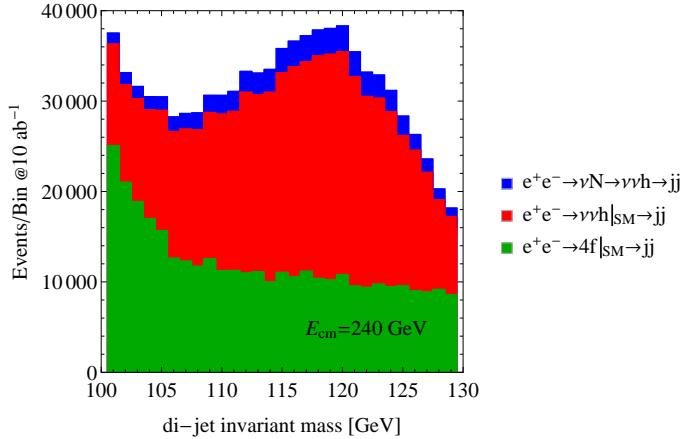}\end{center}
\caption{The di-jet invariant mass spectrum in the mono-Higgs search channel ($jj$ plus missing energy) after simulation of the ILD detector response at $E_{cm}= 240$ GeV. The red and green area denote the dominant and subdominant background, respectively, see text for further details. The blue area denotes the signal from a heavy neutrino with a mass of 152 GeV, and a Yukawa coupling to the electron flavour, $y_{\nu_e}$, saturating the present upper bounds from precision data \cite{Antusch:2014woa,Antusch:2015mia}.}
\label{fig:reconstructed_distributions}
\end{figure}

For illustrative purposes we show the di-jet invariant mass distribution in the mono-Higgs search channel in fig.\ \ref{fig:reconstructed_distributions}. Therein the center-of-mass energy is set to 240 GeV and the model parameters are set to $M= 152$ GeV and $|y_{\nu_e}| = 0.036$, which saturates the present upper bounds at $1\sigma$.
The dominant and subdominant background is represented by the red and green area, respectively. The figure shows how resonant mono-Higgs production, represented by the blue area, contributes substantially to the SM predicted number of events. 

\subsubsection{Kinematic cuts}
For the analysis in the following, we first select the mono-Higgs search channel by applying the above defined selection criteria. After this pre-selection we study the kinematic distributions of the di-jet momentum ($P_{jj}$), the missing transverse momentum ($\cancel{E}_T$), the angular separation of the two jets, and the momentum and energy of the individual jets. We find that the most efficient observable to enhance the significance of the signal, cf.\ eq.~\eqref{eq:significance}, is given by $P_{jj}$. Furthermore, the $\cancel{E}_T$ is very powerful in removing the non-mono-Higgs SM background at $\sqrt{s}=240$ GeV. A detailed list of the applied cuts and the resulting efficiencies can be found in tabs.\ \ref{tab:cuts240}, \ref{tab:cuts350} and \ref{tab:cuts500} in the Appendix.

A comment on b-tagging is in order at this point. We find that, with a nominal selection efficiency of $\sim 0.7$ for a b-flavoured heavy jet, the resulting sensitivity at 240 GeV does not improve the one derived from blindly accepting all hadronic jets. We therefore neglect b-tagging, which may become relevant when a more sophisticated kinematic analysis is applied.

\subsubsection{Future lepton collider sensitivity to the active-sterile mixing parameters}
To establish the sensitivity of the mono-Higgs search channel at future lepton colliders to the active-sterile neutrino mixing, we use the definition from section \ref{sec:parton-sensitivity} for a significance of 1$\sigma$. 
In order to enhance the sensitivity, we have employed a series of kinematic cuts that are listed in tabs.\ \ref{tab:cuts240}, \ref{tab:cuts350} and \ref{tab:cuts500} in the Appendix, together with the resulting numbers of signal and background events.
The resulting sensitivities to the neutrino Yukawa coupling $|y_{\nu_e}|$ for the considered center-of-mass energies are shown in fig.~\ref{fig:sensitivity-detector} for several values of the heavy neutrino mass:

\begin{figure}
\begin{minipage}{0.7\textwidth}
\includegraphics[width=\textwidth]{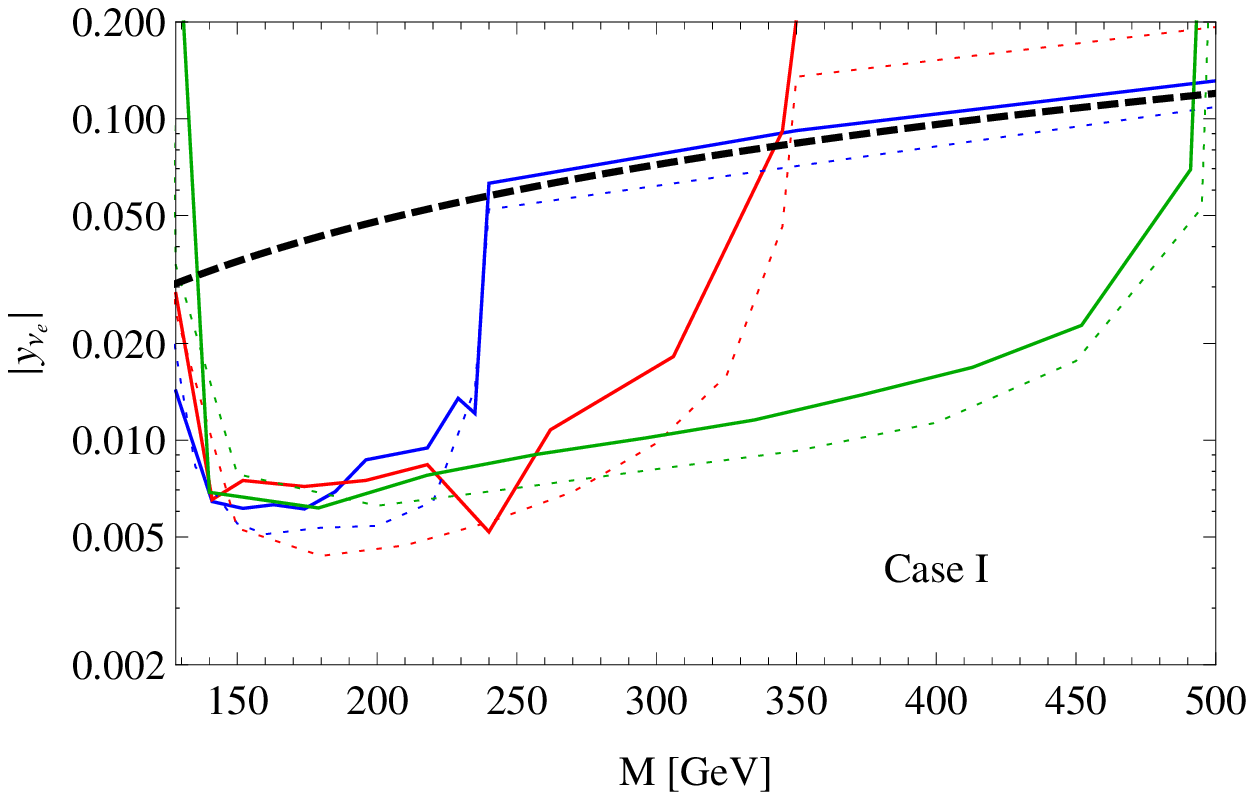}
\end{minipage}
\begin{minipage}{0.29\textwidth}
\begin{center}
\vspace{-20pt}
\includegraphics[width=0.8\textwidth]{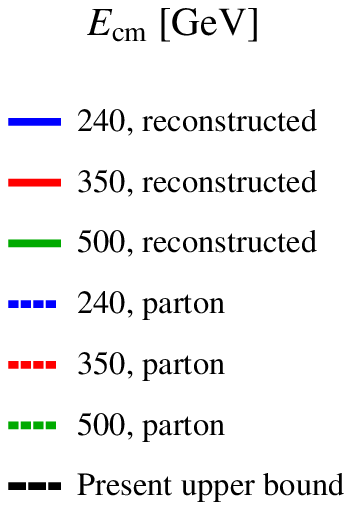}
\end{center}
\end{minipage}
\caption{Future lepton collider sensitivity of the mono-Higgs search channel, i.e.\ di-jet plus missing energy signature, to the neutrino Yukawa coupling $y_{\nu_e}$ at 1$\sigma$. We use the machine performance parameters from tab.~\ref{tab:machines}, simulate the response of the ILD detector and apply kinematic cuts according to tabs.\ \ref{tab:cuts240}, \ref{tab:cuts350} and \ref{tab:cuts500}.}
\label{fig:sensitivity-detector}
\end{figure}

The figure shows that, despite the increased background after reconstruction, the resulting sensitivities from the parton level (denoted by the dotted lines) and the reconstructed level are comparable in magnitude\footnote{It may be possible to further improve on the cuts, up to the point where the sensitivity on the reconstructed level is identical to, or even better than the parton level sensitivity, which we leave for future work.}.
Values of the neutrino Yukawa coupling $y_{\nu_e}$ above the solid lines give rise to a signal that can be distinguished from the SM background with a significance larger than $1\sigma$. The sensitivity at 500 GeV is competitive with 240 and 350 GeV, even for heavy neutrino masses $M\leq 250$ GeV and despite the lower luminosity. Moreover, for values of the heavy neutrino mass above the kinematic threshold (i.e.\ $M>\sqrt{s}$), the signal is due to the non-unitarity effects in mono-Higgs production (cf.\ section~\ref{sec:monoHiggs-NU}), however the corresponding sensitivity is weaker than the present bound.

We remark that resonant mono-Higgs production can give rise to events with larger amount of missing energy compared to the SM. This provides an unambiguous signal without SM background. However, the here considered target luminosities results in ${\cal O}(1)$ and ${\cal O}(10)$ signal events at 240 and 350 GeV, respectively, such that they do not provide an improvement of the sensitivity. We note, that at 500 GeV, the considered luminosity results in less than ${\cal O}(1)$ events of this kind.

Altogether the FCC-ee shows a remarkable sensitivity to the electron neutrino Yukawa coupling which leads to very promising prospects for discovering heavy neutrino signals in the mono-Higgs channel.

\subsubsection{Contamination of SM Higgs-boson parameters}
For the analysis of the Higgs boson at future lepton colliders so-called ``standard cuts'' have been defined\footnote{We thank F. M\"uller for assistance with the ``standard cuts'' for the extraction of mono-Higgs events at lepton colliders.} in \cite{Ono:2013sea}, which we show in tab.~\ref{tab:cuts}. Those cuts are designed to improve the ratio of mono-Higgs events over SM background events. 
However in the case of resonant mono-Higgs production they turn out not to be as efficient in filtering out the additional events from heavy neutrino decays, as is shown in tabs.\ \ref{tab:cuts240} and \ref{tab:cuts350}. 
This contamination of the sample of mono-Higgs events can lead to a deviation from the theory prediction, for instance in the mono-Higgs production cross section, when interpreted in the context of the SM.
We remark that no ``standard cuts'' for 500 GeV exist, and it has only been considered for the FCC-ee very recently. Thus, even though this center-of-mass energy constitutes an excellent environment for studying the mono-Higgs channel, we do not include it in the following.
\begin{table}
\def\arraystretch{1.1 }
\begin{center}
\begin{tabular}{|l|cc|}
\hline
$\sqrt{s}$	&	240 GeV	&	350 GeV \\
\hline
Missing Mass [GeV] 	& 80 $\leq M_{\rm miss} \leq$ 140 & 50 $\leq M_{\rm miss} \leq$ 240 \\
Transverse P [GeV] 	& 20 $\leq P_T \leq$ 70 & 10 $\leq P_T \leq$ 140 \\
Longitudinal P [GeV] & $|P_L|<60$ & $|P_L|<130$ \\ 
Maximum P [GeV]  	& $|P| < 30$ & $|P| < 60$ \\
Di-jet Mass [GeV]	& 100 $\leq M_{jj} \leq$ 130 & 100 $\leq M_{jj} \leq$ 130 \\
Angle (jets) [Rad] 	& $\alpha > $ 1.38 & $\alpha > $ 1.38 \\
\hline
\end{tabular}
\end{center}
\def\arraystretch{1.0}
\caption{``Standard cuts'' from ref.~\cite{Ono:2013sea} to optimize the ratio of mono-Higgs signal to SM background for future lepton colliders. }
\label{tab:cuts}
\end{table}

\begin{figure}
\begin{minipage}{0.6\textwidth}
\includegraphics[width=0.9\textwidth]{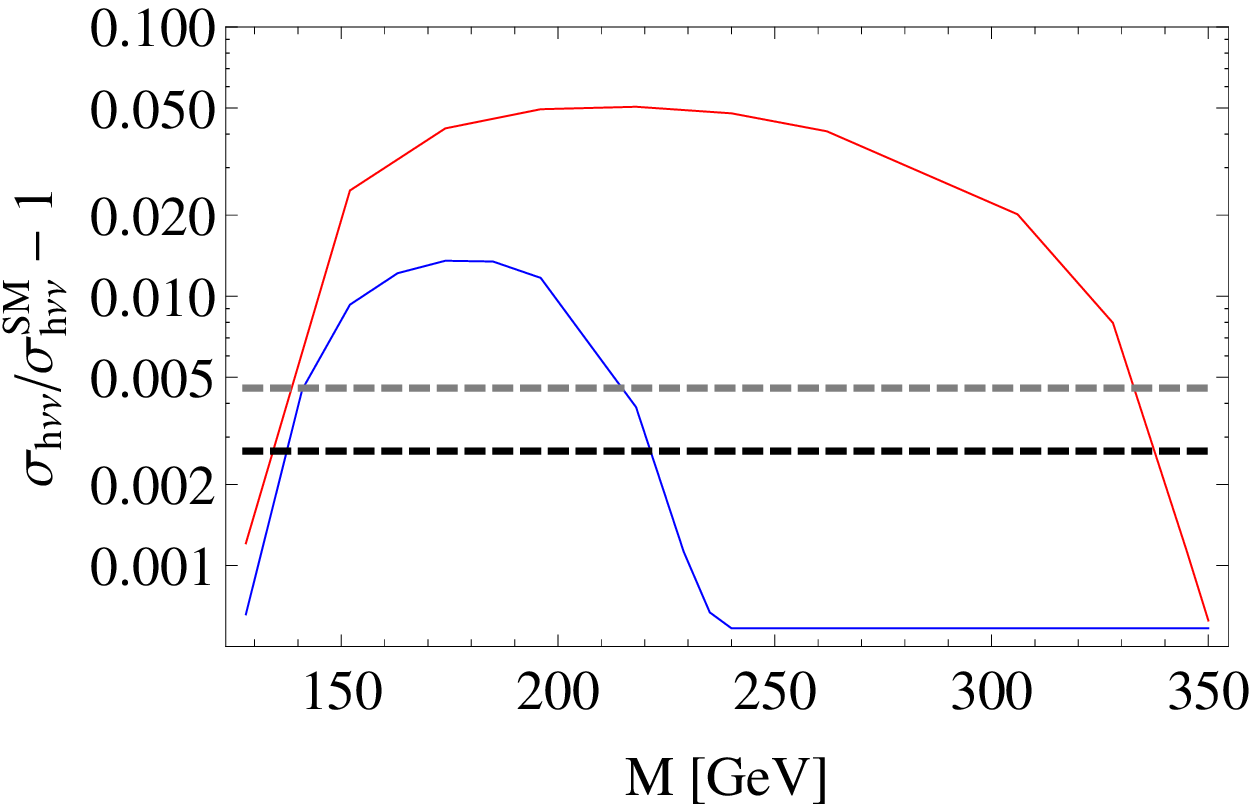}
\end{minipage}
\begin{minipage}{0.35\textwidth}
\includegraphics[width=0.8\textwidth]{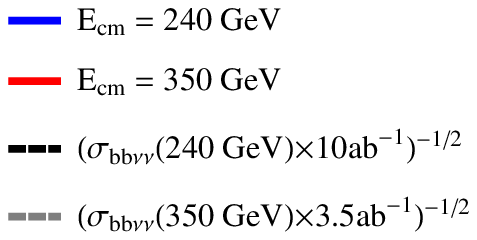}
\end{minipage}
\caption{Heavy neutrino induced deviation of the mono-Higgs production cross section when ``standard cuts'' are applied (cf.\ tab.~\ref{tab:cuts}) to the contaminated event sample when $y_{\nu_e}$ saturates the 1$\sigma$ upper bound.
The horizontal dashed lines denote the relative statistical precision of the SM predicted events $N$, given by $1/\sqrt{N}$.}
\label{fig:contamination}
\end{figure}

In fig.~\ref{fig:contamination} we show the ensuing deviation of the contaminated mono-Higgs production cross section from the SM prediction, for $|y_{\nu_e}|$ saturating the present upper bound. The statistical accuracy at $1\sigma$ are denoted by the black and grey dashed lines for 240 and 350 GeV, respectively.
The figure shows that the deviation of the mono-Higgs production cross section can be significant compared to the experimental accuracy. This can lead to a discrepancy when comparing the Higgs properties derived from the contaminated mono-Higgs sample with the other Higgs channels at 240 GeV \cite{Ruan:2014xxa}.
Moreover, up to a 3$\sigma$ deviation from the SM predicted mono-Higgs production cross section is possible at 240 GeV and at 350 GeV the deviation can be larger than 5$\sigma$.

We emphasise that the shown deviation of the mono-Higgs-production cross section from the SM prediction is fully compatible with present constraints on the active-sterile mixing. Furthermore, if the present non-zero best-fit value for $|\theta_e|$ as reported in refs.\ \cite{Basso:2013jka,Antusch:2014woa,Antusch:2015mia} get confirmed, an observable deviation in the number of mono-Higgs events would be a prediction.

\section{Summary and conclusions}
\label{sec:conclusions}

In this work we have studied Higgs production from sterile neutrinos at future lepton colliders.
We have considered a scenario with a pair of sterile (right-handed) neutrinos that are subject to an approximate ``lepton-number-like'' symmetry. In this scenario the heavy neutrinos (i.e.\ the mass eigenstates) can have masses around the electroweak scale and couple to the Higgs boson with, in principle, unsuppressed Yukawa couplings while accounting for the smallness of the light neutrinos' masses. We refer to this as the ``symmetry protected seesaw scenario'' (SPSS).

The sterile neutrinos in the SPSS enable a novel Higgs production mechanism, given by the on-shell production of a heavy neutrino and its subsequent decay into a light neutrino and a Higgs boson: {\em resonant mono-Higgs production}.
Due to the comparatively large neutrino Yukawa couplings inducing large active-sterile neutrino mixings, the heavy neutrinos can be produced efficiently from lepton collisions. Therefore, future lepton colliders provide a promising environment for testing heavy neutrinos, including signals from resonant mono-Higgs production.

On the contrary, in hadronic collisions the heavy neutrinos can only be produced from the decays of a weak gauge boson, which results in a strong suppression of the heavy-neutrino-production cross section. In addition there are large QCD backgrounds, and only transversal projections of the kinematic observables can be studied. Altogether, the sensitivity to the resonant mono-Higgs production at the LHC is much weaker than at the here considered future lepton colliders (see ref.\ \cite{Basso}).

In order to assess the prospects for testing resonant mono-Higgs production at future lepton colliders, 
we consider the FCC-ee in the following and we expect the results to be representative for the CEPC and indicative for the ILC. 

For $\sqrt{s}$ we consider 240 GeV, 350 GeV and 500 GeV as currently discussed in the working groups.
We have generated Monte Carlo event samples for the SM background and the heavy neutrino signal, where we used the present 1$\sigma$ upper bounds for the active-sterile mixing parameters and we simulated the detector response.

We find that the number of resonantly produced mono-Higgs events can be as large as $\sim 10\%$ of the SM predicted number at 240 GeV. Furthermore, the upper bound on the resonant mono-Higgs events can be up to $\sim 30\%$ and $\sim 40\%$ at $\sqrt{s}$ of 350 and 500 GeV, respectively. Via the mono-Higgs channel, the FCC-ee would be sensitive to the neutrino Yukawa coupling $|y_{\nu_e}|$ (respectively to the active-sterile mixing parameter $|\theta_e|$) down to $\sim 5 \times 10^{-3}$ (cf.\ fig.\ 8). Interestingly, higher $\sqrt{s}$ not only allows for an increased range of testable heavy neutrino masses $M$, but also the signal-to-background ratio increases such that a comparable sensitivity can be achieved with less integrated luminosity. 

Moreover, we have shown that the resonantly produced mono-Higgs events can effectively contaminate the SM analysis of the mono-Higgs channel, as shown in fig.\ 9. With $|\theta_e|$ within the present $1\sigma$ upper bounds, this can lead to deviations from the SM prediction at the percent level, much larger than the estimated future accuracy.

In summary, we discussed a novel channel for Higgs production, namely resonant mono-Higgs production from sterile neutrinos. It can induce sizeable deviations from the SM mono-Higgs prediction and can be used as a sensitive probe of sterile neutrino properties at lepton colliders.

\subsection*{Acknowledgements}
This work has been supported by the Swiss National Science Foundation. We thank Lorenzo Basso for assisting with Madgraph5\_aMC$@$NLO and madanalysis5, for valuable discussions, and for reading the manuscript. We furthermore thank J\"urgen Reuter for assisting with WHIZARD and Michele Selvaggi for helping with the ILD card for DELPHES. O.F. is thankful for stimulating discussions at the first FCC-ee mini-Higgs-Workshop at CERN.

\begin{appendix}
\section*{Appendix: cross sections and cuts}

\begin{table}[!htbp]
\def\arraystretch{1.1 }
\begin{center}
\begin{tabular}{|l||c|c|c|}
\hline
Final state & $\sigma^{\rm SM}@240$ GeV [fb] & $\sigma^{\rm SM}@350$ GeV [fb] & $\sigma^{\rm SM}@500$ GeV [fb] \\
\hline\hline 
$b\bar b$$\nu\nu$	&	146.492	&	134.614	&	183.594	\\
$c\bar c$$\nu\nu$	&	88.0172	&	73.7956	&	82.7041	\\
$jj \nu \nu$		&	528.8	&	463.1	&	500.3	\\
$b\bar b b\bar b$	&	81.2629	&	47.6152	&	25.5571	\\
$b\bar b c\bar c$	&	146.566	&	87.6518	&	51.6446	\\
$b\bar b jj$		&	6820.6	&	4259.5	&	2537.8	\\
$b\bar b e^+ e^-$	&	2080.87	&	2500.82	&	2920.9	\\
$b\bar b \tau^+\tau^-$	&	34.1905	&	19.7975	&	11.0619	\\
$c\bar c \tau^+\tau^-$	&	25.2553	&	15.0695	&	9.15227	\\
$j j \tau^+\tau^-$	&	116.0	&	72.4	&	37.6	\\
$\tau^+\tau^-\nu\nu$	&	235.89	&	163.851	&	119.989	\\
single top		&	0.012	&	63.3	&	1092 \\
$t \bar t$  		&	---	&	322.	&	574. \\
\hline
\end{tabular}
\end{center}
\caption{Included Standard Model four fermion background to the mono-Higgs channel, for details see text. We separated hadronic jets from heavy (charm and bottom) quarks, and denote events with light jets from up, down, strange quarks, and gluons with a $j$.
The cross sections for both tables have been evaluated with WHIZARD 2.2.7 \cite{Kilian:2007gr,Moretti:2001zz}, the efficiency was obtained with the cuts from tab.~\ref{tab:cuts} via madanalysis5 \cite{Conte:2012fm}.}
\label{tab:processtableSM}
\end{table}

\begin{table}[!htbp]
\begin{center} %Center-of-mass energy: 240 GeV
\begin{tabular}{|c||c|c|c|c|c|c|c|c|c|c|}
\hline 
$M$ [GeV]& $P_{jj}$ [GeV] &$N_S$ & $N_B$ & $N_S^{\rm SM}$ \\ 
\hline\hline
128 & $>$ 80 & 308	&	4287	&	25.1	\\
141 & $>$ 70 & 3780	&	18627	&	1327	\\
152 & $>$ 70 & 4846	&	18627	&	2951	\\
163 & $>$ 70 & 5286	&	18627	&	3924	\\
174 & $>$ 60 & 8759	&	34946	&	4387	\\
185 & $>$ 70 & 5652	&	18627	&	4358	\\
196 & $>$ 80 & 1935	&	4287	&	3762	\\
218 & $>$ 70 & 4192	&	18637	&	1113	\\
229 & $>$ 75 & 1505	&	8147	&	182	\\
235 & $>$ 75 & 1966	&	8147	&	29	\\
\hline
\end{tabular}
\end{center}
\caption{List of kinematic cuts for the extraction of the sensitivity in fig.\ \ref{fig:sensitivity-detector}. For all benchmark points for the heavy neutrino mass $M$ at $\sqrt{s}=240$ GeV, the cuts 110 GeV $\leq M_{jj} \leq$ 125 GeV and $\cancel{E}_T > 15$ GeV have been applied. The number of SM background events, $N_B^{\rm SM}$, after application of the ``standard cuts'' is 338600.}
\label{tab:cuts240}
\end{table}

\begin{table}[!htbp]
\begin{center} %Center-of-mass energy: 350 GeV
\begin{tabular}{|c||c|c|c|c|c|c|c|}
\hline 
$M$ [GeV]	& $M_{jj}$ [GeV] & $P_{jj}$ [GeV] &$\cancel{E}_T$ [GeV] &$N_S$ & $N_B$ & $N_S^{\rm SM}$ \\ 
\hline
\hline
128 	& 100 - 130 &100 - 170 & --- 	  & 384	&	109908	&	210	\\
141 	& 110 - 125 & 70 - 160 & --- 	  & 3581	&	17695	&	8652	\\
152 	& 110 - 125 & 80 - 160 & 20 - 100 & 6991	&	86650	&	14874	\\
174 	& 110 - 125 & 50 - 150 & 20 - 100 & 11800	&	120975	&	17562	\\
196 	& 100 - 130 & 50 - 150 & 20 - 100 & 16331	&	171483	&	17937	\\
218 	& 100 - 130 & 50 - 150 & 20 - 100 & 16113	&	171483	&	16948	\\
240 	& 100 - 130 & 50 - 150 & 50 - 100 & 15009	&	14656	&	14504	\\
262 	& 100 - 130 & 70 - 150 & 60 - 100 & 12151	&	126722	&	7016	\\
306 	& 100 - 130 & 110 - 150& 50 - 150 & 6529	&	160592	&	2636	\\
345	& 100 - 130 & 120 - 160& 20 - 150 & 331	&	163809	&	183	\\
\hline
\end{tabular}
\end{center}
\caption{List of kinematic cuts for the extraction of the sensitivity in fig.\ \ref{fig:sensitivity-detector} for $\sqrt{s}=350$ GeV. The number of SM background events, $N_B^{\rm SM}$, after application of the ``standard cuts'' is 359500.}
\label{tab:cuts350}
\end{table}

\begin{table}[!htbp]
\begin{center} %Center-of-mass energy: 500 GeV
\begin{tabular}{|c||c|c|c|c|}
\hline 
$M$ [GeV]	& $P_{jj}$ [GeV] & $\cancel{E}_T$ [GeV] & $N_S$ & $N_B$ \\
\hline\hline 
140 & $>$ 170 	& $<$ 100 & 6248	&	7550	\\
179 & $>$ 100 	& $<$ 100 & 25176	&	29453	\\
218 & --- 	& --- 	&   43304	&	101672	\\
257 & --- 	& --- 	&   44691	&	101672	\\
296 & --- 	& 50 - 200 &37571	&	65326	\\
335 & --- 	& 70 - 180 &30710	&	44572	\\
374 & --- 	& 90 - 180 &21766	&	29854	\\
413 & 160 - 220 & --- 	&   14926	&	20541	\\
452 & 170 - 230 & --- 	&   8551	&	15643	\\
495 & $>$ 220 	& --- 	&   845	&	9533	\\
\hline
\end{tabular}
\end{center}
\caption{List of kinematic cuts for the extraction of the sensitivity in fig.\ \ref{fig:sensitivity-detector}. For all benchmark points for the heavy neutrino mass $M$ at $\sqrt{s}=500$ GeV, the pre-selection cuts have been slightly loosened to 100 GeV $\leq M_{jj} \leq$ 140 GeV.}
\label{tab:cuts500}
\end{table}

\end{appendix}
\newpage

\bibliographystyle{unsrt}

\end{document}